\documentclass[11pt]{article}
\usepackage{amsmath}
\usepackage{amssymb}

\usepackage{scicite}

\usepackage{subfigure}
\usepackage{times}
\usepackage{epsfig}
\usepackage{url}

\newtheorem{definition}{Definition}[section]
\newtheorem{lemma}{Lemma}[section]
\newtheorem{theorem}{Theorem}[section]

\newtheorem{proof}{Proof}

\newenvironment{sciabstract}{%
\begin{quote} \bf}
{\end{quote}}

\newcounter{lastnote}

\begin{document}


\title{ Community Structures Are Definable in Networks: A Structural Theory of Networks
\footnote{State Key Laboratory of Computer Science, Institute of Software,
Chinese Academy of Sciences, P. O. Box 8718, Beijing, 100190, P. R.
China.  Email: \{angsheng, yicheng, lijk\}@ios.ac.cn.
Correspondence: \{angsheng, yicheng\}@ios.ac.cn.
\newline
Angsheng Li is partially supported by the Hundred-Talent Program of
the Chinese Academy of Sciences. All authors are supported by the
Grand Project ``Network Algorithms and Digital Information" of the
Institute of software, Chinese Academy of Sciences, and NSFC grant
No. 61161130530.}}

\author{ Angsheng Li$^{1}$,  Yicheng Pan$^{1,3 }$, Jiankou Li$^{1,2}$ \\
\normalsize{$^{1}$State Key Laboratory of Computer Science}\\
\normalsize{ Institute of Software, Chinese Academy of Sciences}\\
\normalsize{$^{2}$University of Chinese Academy of Sciences,
P. R. China}\\
\normalsize{$^{3}$State Key Laboratory of Information Security}\\
\normalsize{ Institute of Information Engineering, Chinese Academy of Sciences,
P. R. China} }


\date{}



\baselineskip24pt

\maketitle

\begin{sciabstract}

 Community detecting is one of the main approaches to understanding
network structures. However it has been a longstanding challenge to give a definition for community structures of networks.
We found that neither randomness in the ER model nor the preferential attachment in the PA model
is the mechanism of community structures of networks, that community structures are universal in real networks, that
community structures are definable in networks, that communities are interpretable in networks, and that homophyly is the mechanism of community structures and a structural theory of networks.
We proposed the notions of entropy- and conductance-community structures. It was shown that the two
definitions of the entropy- and conductance-community
structures and the notion of modularity proposed by physicists are all equivalent in defining community structures of
networks, that neither randomness in
the ER model nor preferential attachment in the PA model is the mechanism of community structures of
networks, that there is an empirical criterion for deciding the
existence and quality of community structures in networks, and that the existence of community structures is a
universal phenomenon in real networks. This poses a fundamental question: What are the mechanisms of community
structures of real networks? To answer this question, we proposed a homophyly model of networks.
It was shown that networks of our model satisfy a series of new topological, probabilistic and
combinatorial principles, including a fundamental principle, a community structure principle,
a degree priority principle, a widths principle, an inclusion and infection principle, a king node principle and a predicting principle etc. The new principles
provide a firm foundation for a structural theory of networks. Our homophyly model demonstrates that
homophyly is the underlying mechanism of community structures of networks, that
nodes of the same community share common features, that power law and small world property are never
obstacles of the existence of community structures in networks, that community structures are {\it definable} in networks, and that (natural) communities are
{\it interpretable}. Our theory
provides a foundation for analyzing the properties and
roles of community structures in robustness, security, stability, evolutionary games,
predicting and controlling of networks.

\end{sciabstract}

\section{Background}

Defining community structures in networks is a fundamental challenge
in modern network theory. Our notions of the entropy- and
conductance-community structures are information theoretical and
mathematical definitions respectively. We found that our definitions of
entropy-, conductance- and the modularity-community structures \cite{NG2004} are
equivalent in defining community structures of networks, that randomness and preferential attachment are not mechanisms
of community structures of networks, and that
community structures are universal in real networks. This shows that
community structure is a phenomenon definable in networks.
The discoveries here provide a foundation for a new theory of community (or local)
structures of networks.

Generally speaking, the missing of a structural theory of networks hinders us from rigorous analysis of networks and networking data. In fact,
the current state of the art shows that the current tools for analyzing networks and networking data are mainly the probabilistic or statistical methods, which
of course neglect the structures of data.

However, structures are essential. In nature and society, especially in the current highly connected world, we
observe that mechanisms determine the structures, and that structures determine the
properties, which could be a new hypothesis of real network data. In particular, a grand challenge in network science is apparently the missing of a structural theory of networks.
Community structures determine the local properties of
networks that may have global implications.

Network  has become a universal topology in science, industry,
nature and society. Most real networks satisfy a power law degree
distribution~\cite{Bar1999}, \cite{Bar2009}, and a small world
phenomenon~\cite{M1967}, \cite{WS1998}, \cite{K2000}.

Community detecting or clustering is a powerful tool for
understanding the structures of networks, which has already been extensively
studied \cite{CR2010}, \cite{CA2005}, \cite{RCCLP2004},
\cite{CNM2004}, \cite{NEW2004}. Many definitions of communities have
been introduced, and various methods for detecting communities have been
developed in the literature, see \cite{For2010} for a recent survey.
However, the problem is still very hard, not yet satisfactorily
solved. The approach of community finding takes for granted that
networks have community structures. The fundamental questions are
thus: Are communities objects naturally formed in a network or
simply outputs of a graphic algorithm? Can we really take for
granted that networks have community structures? Are community
structures of networks robust? What are the natural mechanisms which
generate the community structure of a network, if any? What can we
 do with the communities? How can we test the quality of various community finding algorithms?

Our experiments here showed that community structures are definable in networks, that randomness and preferential attachment are not mechanisms of community structures of networks, that there is
an empirical criterion for deciding the existence and quality of community structures of networks, and that the existence of community structures is a universal phenomenon of real world networks.
This predicts that community structures have their own mechanisms other than the well-known ones of classical models of networks.

It is easy to recognize that the key to answer all the questions listed above is to understand the mechanisms of community structures of networks.

We proposed a new model of networks, the {\it homophyly model} below, by introducing the new mechanism of homophyly in
the classical preferential attachment model. Our model constructs networks dynamically by using the natural mechanisms of
preferential attachment and homophyly.

Our homophyly model explores that homophyly is the
mechanism of community structures in networks, that community structures are provably definable in networks, and that communities are interpretable in networks.
It was shown that the homophyly networks satisfy simultaneously
a series of new topological, probabilistic and combinatorial principles, including a fundamental principle, a
community structure principle, a degree priority principle, a widths principle, an inclusion and infection principle, a king node principle and a predicting principle etc.
The new principles we found here provide a foundation for analyzing the properties and roles of community structures in new issues of networks
such as robustness, security, stability of networks, evolutionary game and predicting in networks, and controlling of networks. Our results here provide a firm first step
for us to develop a structural theory of networks, which would be essential to many important new issues of networks and networking data.

\section{Definitions of Community Structures of Networks} \label{sec:defnew}

The first definition of community structures is the notion of modularity.
Newman and Girvan \cite{NG2004} defined the notion of modularity to
quantitatively measure the quality of community structure of a
network. It is built based on the assumptions that random graphs are
not expected to have community structure and that a network has a
community structure, if it is far from random graphs.

Let $G=(V,E)$ be a graph with $n$ nodes and $m$ edges and
$\mathcal{P}$ be a partition of nodes in $V$. The {\it  modularity
of $G$ by $\mathcal{P}$} is defined by

$$\sigma^{\mathcal{P}}(G)=\frac{1}{2m} \sum\limits_{i,j} (A_{ij}-P_{ij})\delta (C_i, C_j),$$

\noindent where the sum runs over all pairs of vertices, $A$ is the
adjacency matrix, $P_{ij}$ is the expected number of edges between
nodes $i$ and $j$ in a null graph, i.e., a random version of $G$.
$\delta (C_j,C_j)=1$ if $C_i=C_j$, and $0$ otherwise, $C_k$ is an
element of the partition $\mathcal{P}$.

A standard null model imposes that the expected degree after
averaging over all possible configurations matches the actual degree
of the original graph \cite{F09}. Such a null model is essentially
equivalent to the configuration model \cite{CL06}, in which each node
$i$ is associated with $d_i$ half-edges, where $d_i$ is the degree
of node $i$ in $G$, and all the half-edges are joined randomly. It
is easy to obtain that $P_{ij}=d_i d_j/2m$ and the modularity of
$G$ by $\mathcal{P}$ can then be rewritten as
\begin{equation} \label{eqn:modularity}
\sigma^{\mathcal{P}}(G)=\sum\limits_{l=1}^L \left[
\frac{k_l}{m}-\left(\frac{V_l}{2m}\right)^2 \right],
\end{equation}
where $L$ is the number of modules in partition $\mathcal{P}$, $k_l$
is the number of edges whose both ends are in module $l$, and $V_l$
is the sum of the degrees of the nodes in $l$, also called the {\it
volume} of $l$. Note that the first term of each summation represents
the fraction of edges of $G$ inside the module and the second term
represents the expected fraction of edges that would be in the
null model.

We define the {\it modularity of $G$} as

$$\sigma(G)=\max_{\mathcal{P}}\{\sigma^{\mathcal{P}}(G)\}.$$

$\sigma (G)$ is a real number in $[0,1]$. The larger $\sigma (G)$
is, the better community structure $G$ has. We define the {\it
modularity community structure ratio } (M-community structure ratio)
of $G$ to be the modularity of $G$.

The second definition is based on random walks.
The intuition is that random walks from a node in a quality community are not easy
to go out of the community. We define the notion of entropy community structure ratio of a network.
 We consider the entropy of a network and of a network given by a partition of nodes of the network.

Let $G=(V,E)$ be a graph with $n$ nodes and $m$ edges, and
$\mathcal{P}$ be a partition of $V$. We use $L^U(G)$ to denote the
minimum average number of bits to represent a single step of random walk
(in the stationary distribution) with a uniform code in $G$, and
$L^{\mathcal{P}}(G)$ to denote the minimum average number of bits to
represent that with an aforementioned ``module-node" code.

 By
information theoretical principle,
\begin{equation} \label{eqn:entropy_G}
L^U(G)=-\sum\limits_{i=1}^n \frac{d_i}{2m} \cdot
\log_2\frac{d_i}{2m},
\end{equation}
where $d_i$ is the degree of node $i$.

\begin{equation} \label{eqn:entropy_partition}
L^{\mathcal{P}}(G)=-\sum\limits_{j=1}^L \sum\limits_{i=1}^{n_j}
\frac{d_i^{(j)}}{2m} \cdot \log_2\frac{d_i^{(j)}}{V_j}-\frac{m_g}{m}
\left( \sum\limits_{j=1}^L \frac{V_j}{2m} \cdot \log_2\frac{V_j}{2m}
\right),
\end{equation}
where $L$ is the number of modules in partition $\mathcal{P}$, $n_j$
is the number of nodes in module $j$, $d_i^{(j)}$ is the degree of
node $i$ in module $j$, $V_j$ is the volume of module $j$, and $m_g$
is the number of edges crossing two different modules.

We define the {\it entropy community structure ratio of $G$ by
$\mathcal{P}$} by

$$\tau^{\mathcal{P}} (G) = 1 - \frac{L^{\mathcal{P}}(G)}{ L^U(G)}.$$

We define the {\it entropy community structure ratio of $G$}
(E-community structure ratio of $G$) by

$$\tau (G)=\max_{\mathcal{P}}\{\tau^{\mathcal{P}}(G)\}.$$

We notice that similar idea has been used in community detecting, for instance, Rosvall and Bergstrom \cite{RB2008}
proposed an algorithm to detect communities by compressing a description of the information flow by using the Huffman code to encode prefix-freely each module and each
node. Our definition here is purely an information theoretical notion.

Both definitions of the modularity and entropy community structure
ratio depend on partitions of $G$. In these definitions, the
existence of a community structure of a graph means that there is a
``good partition" for the graph. However the intuitive relationship
between ``good partitions" and community structures is not clear,
and more seriously, both the two definitions are not convenient for
us to compare the community finding algorithms, which usually do
not give partitions of networks. In fact, a community finding
algorithm may find overlapping communities, and may neglect part of
the nodes in the community findings. In addition, both modularity and the entropy community structure ratio can be
regarded as global definitions of community structures of networks.

We will introduce a mathematical definition based on conductance, which is
applicable to overlapping communities, and to partial solutions of
searching etc. We define the {\it conductance community structure ratio of a network}. To describe our definition, we recall the notion of conductance.

Given a graph $G=(V,E)$, and a set $S\subset V$, the conductance of
$S$ in $G$ is defined by

$$\Phi^G(S)=\frac{|E(S,\bar{S})|}{\min\{{\rm vol}(S), {\rm
vol}(\bar{S})\}},$$

\noindent where $\bar{S}$ is the complement of $S$ in $G$,
$E(S,\bar{S})$ is the set of all edges with one endpoint in $S$ and
the other in $\bar{S}$, ${\rm vol}(X)$ is the volume of $X$.

Clearly, a community of a network must satisfy certain basic
conditions, for instance: the induced subgraph must be connected, and
the size of the community is not too small, and not too large. For
this, we define:

\begin{definition} \label{def:posscommunity} (Possible community)
Given a graph $G=(V,E)$, let $n=|V|$, and $S\subset V$. We say that
$S$ is a {\it possible community} of $G$, if:

\begin{enumerate}
\item [(1)] The induced subgraph $G_S$ of $S$ in $G$ is connected,

\item [(2)] $|S|\geq\log n$, i.e., the size of $S$ is not too small,
and

\item [(3)] $|S|\leq\sqrt{n}$, that is, the size of $S$ is not too
large.

\end{enumerate}

\end{definition}

\begin{definition} \label{def:communities} (Conductance community
structure ratio of communities) Let $G=(V,E)$ be a graph, and
$n=|V|$. Suppose that $\mathcal{X}=\{ X_1, X_2,\cdots, X_l\}$ is a
set of possible communities of $G$ (overlapping is permitted). Then,

\begin{enumerate}
\item [(1)] Let $X=\cup_jX_j$.
\item [(2)] For each $x\in X$, suppose that $X_{j_1},
X_{j_2},\cdots, X_{j_r}$ are all the possible communities
$X_j\in\mathcal{X}$ that contain $x$, then define

$$
a^{\mathcal{X}}(x) =  \sum\limits_i^r(1-\Phi
(X_{j_i}))/r.
$$

\item [(3)] Define the conductance community structure ratio of $G$ by $\mathcal{X}$ by

$$\theta^{\mathcal{X}}(G)=\frac{1}{n}\cdot \sum\limits_{x\in
X}a^{\mathcal{X}}(x).$$

\end{enumerate}

\end{definition}

By using $\theta^{\mathcal{X}}(G)$, we define the conductance community
structure ratio of a network.

\begin{definition} \label{def:networks} (Defining $\theta (G)$) Let
$G=(V,E)$ be a network. We define the {\it conductance community
structure ratio of $G$} by

$$\theta (G)=\max_{\mathcal{X}}\{\theta^{\mathcal{X}}(G)\}.$$

\end{definition}

The conductance community structure ratio (C-community structure ratio, for short) can be interpreted as a
mathematical definition of community structures of networks. It is
a local definition of community structures of networks.

Suppose that $G=(V,E)$ is an expander with the following properties:
for any nontrivial set $S\subset V$, the conductance $\Phi
(S)>\alpha$ for some large constant $\alpha$. Then by definition,
$\theta (G)<1-\alpha$. Therefore $\theta (G)$ cannot be large for
graph $G$ with expanding property. In particular, for a nontrivial  network $G$
constructed from the PA model, with probability $1-o(1)$, $\theta
(G)<1-\alpha$ for some large constant $\alpha$.

More importantly, the conductance community structure ratio can be defined for algorithms.

\begin{definition} \label{def:algorithm} (Defining $\theta^{\mathcal{A}} (G)$) Let
$G=(V,E)$ be a network. Let $\mathcal{A}$ be a community detecting
algorithm. Suppose that $\mathcal{X}$ is the set of all possible
communities of $G$ found by algorithm $\mathcal{A}$. Then define:

$$\theta^{\mathcal{A}} (G)=\theta^{\mathcal{X}}(G).$$

\end{definition}

By using Definition \ref{def:algorithm}, we can define a community
finding problem as follows.

\begin{definition} \label{def:problem} (Community finding problem)
To design an algorithm, $\mathcal{A}$ say, such that for any network
$G$, $\theta^{\mathcal{A}}(G)$ is maximized.

\end{definition}

This gives rise to the algorithmic problem of searching.

\section{Community Structures Are Robust} \label{sec:robust}

Now we have three definitions of community structure ratios of
networks. Are there any relationships among the three definitions
for the quality of community structures of networks, i.e., the M-,
E-, and C-community structure ratios? Do the three
definitions give the same answer to the question whether or not a
network has a community structure? We conjecture that the answer is
yes. For this, we propose the following:

 {\it Community structure hypothesis}: Given a network $G$, the following properties are
equivalent,

\begin{enumerate}

\item [1)] $G$ has an M-community structure,

\item [2)] $G$ has an E-community structure, and

\item [3)] $G$ has a C-community structure.

\end{enumerate}

We verify the community structure hypothesis by computing the M-,
E-, and C-community structure ratios for networks of
classical models. The first model is the ER model \cite{ER1960}. In
this model, we construct graph as follows: Given $n$ nodes
$1,2,\cdots, n$, and a number $p$, for any pair $i, j$ of nodes $i$
and $j$, we create an edge $(i,j)$ with probability $p$. The second
is the PA model \cite{Bar1999}. In this model, we construct a
network by steps as follows: At step $0$, choose an initial graph
$G_0$. At step $t>0$, we create a new node, $v$ say, and create $d$
edges from $v$ to nodes in $G_{t-1}$, chosen with probability
proportional to the degrees of nodes in $G_{t-1}$, where $G_{t-1}$ is the
graph constructed at the end of step $t-1$, and $d$ is a natural
number.

\begin{figure}
  \centering

    \includegraphics[width=3in]{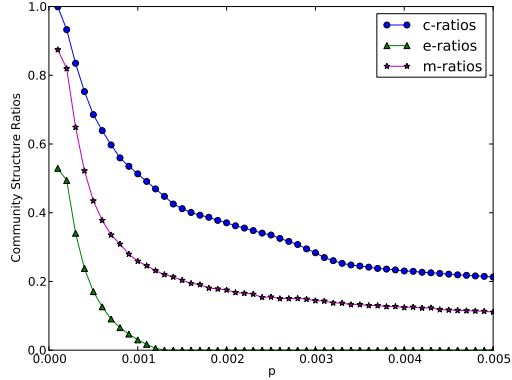}

  \caption{{This figure gives the E-, M- and C-community structure ratios
  (denoted by e-, m- and c-ratios respectively) of networks, for $n=10,000$, and for $p$ up to $0.005$ of the ER model.}}
\label{figure_er_modularity}
\end{figure}

\begin{figure}
  \centering
  \includegraphics[width=3in]{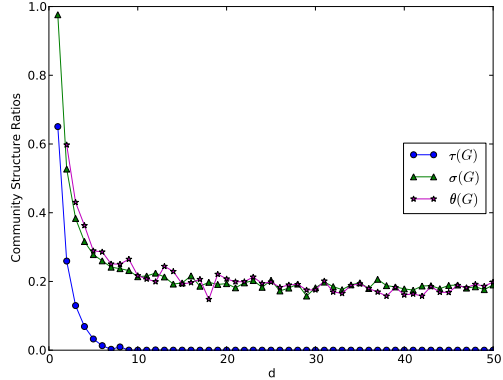}
  \caption{{This figure gives the E-, M- and C-community structure ratios
  (denoted by e-, m- and c-ratios respectively) of networks, for $n=10,000$, and for $d\leq 50$ of the PA model.}}\label{figure_ba_modularity}
\end{figure}

We depict the curves of the M-, E-, and C-community structure
ratios of networks of the ER model and the PA model in Figures
~\ref{figure_er_modularity} and ~\ref{figure_ba_modularity}
respectively.

From Figures \ref{figure_er_modularity} and
\ref{figure_ba_modularity}, we observe that the following results
hold:

\begin{enumerate}
\item [(1)] The curves of the M-, E-, and C-community structure ratios of
networks generated from the ER model are similar.
\item [(2)] The curves of the M-, E-, and C-community structure
ratios of networks generated from the PA model are similar.
\item [(3)] Nontrivial networks of the ER and PA models fail to have
a community structure.
\item [(4)] For a network constructed from either the ER or the PA
model, if the average number of edges $d$ of the network is bounded
by a small constant, $5$ say, then the network has some community
structure.

\end{enumerate}

(1) and (2) show that the community structure hypothesis holds for
all networks generated from the classic ER and PA models. We notice
that every network essentially uses the mechanisms of both the ER
and the PA models. Our results here imply that the hypothesis may
hold for all networks. (3) and (4) show that neither randomness in
the ER model nor preferential attachment in the PA model alone is
the mechanism of community structures of networks.

\section{Community Structures Are Universal in Real Networks}
\label{sec:real}

By observing the experiments in Figures~\ref{figure_er_modularity}
and ~\ref{figure_ba_modularity}, we have that for a network $G$ of
either the ER model or the PA model, the following three properties
(1), (2) and (3) either  hold simultaneously or fail to hold
simultaneously:

\begin{enumerate}
\item [(1)] the E-community structure ratio of $G$, $\tau (G)$, is
greater than $0$,

\item [(2)] the M-community structure ratio of $G$, $\sigma (G)$, is
greater than $0.3$, and

\item [(3)] the C-community structure ratio of $G$, $\theta (G)$, is
greater than $0.3$.
\end{enumerate}

This result suggests an {\it empirical criterion} for deciding
whether or not a network has a community structure. Let $G$ be a network, then

\begin{enumerate}
\item We say that $G$ has a community structure if the E-, M-, and C-community structure ratios of $G$
are greater than $0$, $0.3$ and $0.3$ respectively.

\item The values $\sigma (G)$, $\tau (G)$ and $\theta (G)$ represent
the quality of community structure of $G$, the larger they are, the
better community structure $G$ has.

\end{enumerate}

By using the empirical criterion of community structure of networks,
we are able to decide whether or not a given network has a
community structure.

We implemented the experiments of the M-, E- and C-community
structure ratios for $22$ real networks, which are given in
Table~\ref{table_statistic}. From  the table, we have that if one of
the M-, E- and C-community structure ratios is high, then the other
two ratios are high too, and that for each of the networks, the E-,
M- and C-community structure ratios are greater than $0$,
$0.3$ and $0.3$ respectively.

\begin{table}
 \centering
\begin{tabular} {|c|c|c|c|}
\hline
network &$\tau (G)$&$\sigma (G)$&$\theta (G)$\\
\hline
cit\_hepph&0.22&0.56&0.37\\
\hline
cit\_hepth&0.2&0.53&0.36\\
\hline
col\_astroph&0.24&0.51&0.49\\
\hline
col\_condmat&0.37&0.64&0.76\\
\hline
col\_grqc&0.44&0.79&0.89\\
\hline
col\_hepph&0.26&0.58&0.7\\
\hline
col\_hepth&0.39&0.69&0.83\\
\hline
email\_enron&0.21&0.5&0.63\\
\hline
email\_euall&0.39&0.73&0.76\\
\hline
p2p4&0.11&0.38&0.36\\
\hline
p2p5&0.11&0.4&0.36\\
\hline
p2p6&0.12&0.39&0.38\\
\hline
p2p8&0.15&0.46&0.46\\
\hline
p2p9&0.15&0.46&0.42\\
\hline
p2p24&0.21&0.47&0.48\\
\hline
p2p25&0.23&0.49&0.5\\
\hline
p2p30&0.24&0.5&0.53\\
\hline
p2p31&0.25&0.5&0.52\\
\hline
roadnet\_ca&0.67&0.99&0.98\\
\hline
roadnet\_pa&0.66&0.99&0.98\\
\hline
roadnet\_tx&0.67&0.99&0.98\\
\hline
\end{tabular}
 \caption{The entropy, modularity and
conductance community structure ratios of real networks, written by
$\tau (G)$, $\sigma (G)$ and $\theta (G)$ respectively.}
\label{table_statistic}
\end{table}

The experiments in Table~\ref{table_statistic} show that the
community structure hypothesis holds for (each of the) real
networks, which further validates the community structure hypothesis,
and that the existence of community structure is a universal phenomenon for (almost all) real
networks.

By observing the experiments in Figures ~\ref{figure_er_modularity}
and ~\ref{figure_ba_modularity}, and Table~\ref{table_statistic}, we
have that the community structure of a network is independent of
which of the three definitions of  community structures, i.e.,
the E-, M- and C-community structures, is used. This shows
that community structures are robust in networks, and that the
existence of community structures in real networks is a universal
phenomenon, independent of both definitions of community structures and algorithms for finding the communities.

By observing all the curves in Figures~\ref{figure_er_modularity}
and ~\ref{figure_ba_modularity}, and all experiments
in Table 1 again, our conclusions are further validated. That is:

\begin{enumerate}
\item [(1)] Community structures are robust and hence definable in networks.
\item [(2)] Community structures are universal in real networks.
\item [(3)] Neither randomness nor preferential attachment is the mechanism
of community structures of networks.

\end{enumerate}

(1) implies that community structures can be theoretically analyzed
in networks, and that community structures are objective existence
in networks, instead of simply outputs of algorithms. This suggests
a fundamental issue to investigate the role of community structures
of networks. (2) and (3) suggest some fundamental questions such as:
what are the mechanisms of community structures of real networks?
What roles do the community structures play in real networks?

\section{Homophyly Networks and Theorems} \label{sec:homomodel}

Recall that community structures are definable in networks and
universal in real networks. Real networks are from a
wide range of disciplines of both social and physical sciences. This
hints that community structures of real networks may be the result
of natural mechanisms of evolutions of networking systems in nature
and society. Therefore mechanisms of community structures of real
networks must be natural mechanisms in nature and society.

In both nature and society, whenever an individual is born, it will
be different from all the existing individuals, it may have its own
characteristics from the very beginning of its birth. An individual
with different characteristics may develop links to existing
individuals by different mechanisms, for instance, preferential
attachment or homophyly.

We propose our homophyly model based on the above intuition. It
constructs a network dynamically by steps as follows.

\begin{definition} \label{def:homomodel} (Homophyly model) Let $d$
be a natural number, and $a$ be a homophyly exponent. The homophyly
model constructs networks by steps.

\begin{enumerate}
\item [(1)] Let $G_2$ be an initial graph with two nodes connected by $d$ multi-edges.
Each node is associated with a distinct color and called seed.

For $i+1>d$, let $G_i$ be the graph constructed at the end of step
$i$. Let $p_{i}=\frac{1}{\log^a i}$.

\item [(2)] Create a new node $v$.

\item [(3)] With probability $p_{i}$, $v$ chooses a new color, in which
case, we say that $v$ is a seed node, and create $d$ edges $(v,u_j)$
for $j=1,2,\cdots, d$ such that each $u_j$ is chosen with
probability proportional to the degrees of nodes in $G_i$.
\item [(4)] Otherwise, $v$ chooses an old color, in which case:

\begin{enumerate}
\item $v$ chooses randomly and uniformly an old color as its own
color, and

\item create $d$ edges $(v,u_j)$ for $j=1,2,\cdots, d$ such that
each $u_j$ is chosen with probability proportional to the degrees of
nodes among all the nodes sharing the same color with $v$.

\end{enumerate}

\end{enumerate}

\end{definition}

The homophyly model constructs networks dynamically with both
homophyly and preferential attachment as its mechanisms. It better
reflects the evolution of networking systems in nature and society.
We call the networks constructed from the homophyly model homophyly
networks.

\begin{figure}
  \centering
\includegraphics[width=3in]{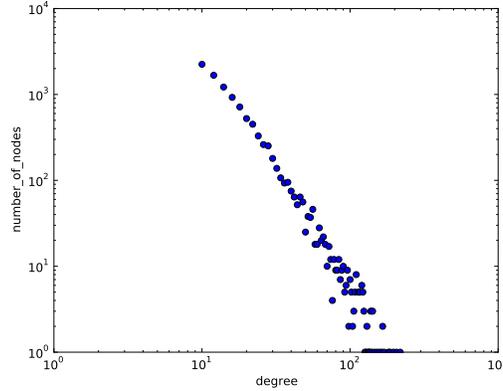}
\caption{Power law distribution of a homophyly network: $n=10,000$,
$a=1.2$ and $d=5$. }
 \label{figure_homophyly_deg_seq}
\end{figure}

\begin{figure}
  \centering
\includegraphics[width=3in]{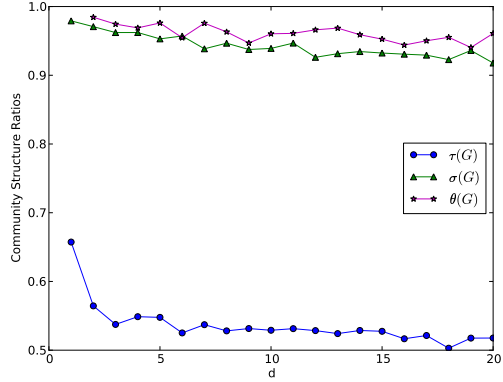}
\caption{The E-, M- and C-community structure ratios (denoted by e-,
m- and c-ratios respectively) of a homophyly network for $n=10,000$
and $a=1.2$.} \label{figure_homophyly}
\end{figure}

We will show that homophyly networks satisfy a series of new
principles, including the well known small world and power law properties.
At first, it is easy to see that the homophyly networks have
the small diameter property, which basically follows from the
classic PA model. Secondly, the networks follow a power law, for
which we see Figure \ref{figure_homophyly_deg_seq} for the intuition. At last,
they have a nice community structure, for which we depict the entropy-,  conductance-community structure ratios, and the
 modularity- \cite{NG2004} of some homophyly networks in Figure \ref{figure_homophyly}.
From Figure \ref{figure_homophyly}, we know that
the entropy-, modularity- and conductance-community structure ratios
of the homophyly networks are greater than $0.5$, $0.9$ and $0.9$
respectively. Therefore the homophyly networks have a community
structure by our criterion in Section \ref{sec:real}.

We notice that the homophyly model is dynamic with homophyly and preferential attachment as its mechanisms. The
initial graph $G_2$ could be any finite colored graph, which does
not change the statistic characteristics of the model. We use
$\mathcal{H}(n,a,d)$ to denote the set of networks constructed from
the homophyly model with average number of edges $d$ and homophyly
exponent $a$. We call a set of nodes of the same color, $\kappa$
say, a {\it homochromatic set}, written by $S_\kappa$. We say that a
homochromatic set is {\it created at time step $t$}, if the seed node of the set is created at time step $t$.

Here we verify that homophyly networks satisfy a number of new topological, probabilistic
and combinatorial principles, including the fundamental principle, the community structure principle, the degree
 priority principle, the widths principle, the inclusion and infection principle, the king
node principle and the predicting principle below.

At first, we have a fundamental theorem.
\bigskip

\begin{theorem} \label{thm:fundtheorem} (Homophyly theorem)
Let $a>1$ be the homophyly exponent, and $d\geq 4$ be a natural
number. Let $G=(V,E)$ be a network constructed by
$\mathcal{H}(n,a,d)$.

Then with probability $1-o(1)$, the following properties hold:

\begin{enumerate}

\item [(1)] (Basic properties):

\begin{enumerate}

\item [(i)] (Number of seed nodes is large) The number of seed nodes is bounded in the interval $[\frac{n}{2\log^a n},\frac{2n}{\log^a
n}]$.

\item [(ii)] (Communities whose vertices are interpretable by common features are small) Each homochromatic set
 has a size bounded by $O(\log^{a+1} n)$.

\end{enumerate}

\item [(2)] For degree distributions, we have:

\begin{itemize}

\item [(i)] (Internal centrality) The degrees of the induced
subgraph of a homochromatic set follow a power law.

\item [(ii)] The degrees of nodes of a homochromatic set follow a
power law.
\item [(iii)] (Power law) Degrees of nodes in $V$ follow a power law.
\item [(iv)] (Holographic law) The power exponents in (i) - (iii) above are all the same.

\end{itemize}

\item [(3)] For node-to-node distances, we have:
\begin{itemize}

\item [(i)] (Local communication law) The induced subgraph of a homochromatic set has a diameter bounded by $O(\log\log
n)$.

\item [(ii)] (Small world phenomenon) The average node to node distance of $G$ is bounded by $O(\log^2 n)$.
\end{itemize}

\end{enumerate}

\end{theorem}

\smallskip

(1)(i) gives an estimation on the number of communities.
(1)(ii) shows that the induced subgraph of a homochromatic set is a
community in which all the nodes share common features, the same
color here, and that a community interpretable by common features is small. (2)(i)-(2)(iv) show that power law is holographic in
networks of the homophyly model, and that a community has an internal centrality in the sense that it has a small set dominating the community. This predicts that the holographic
property may hold for many real networks, and that natural communities of a real network may have the internal centrality. (3) shows that $G$ have
the small world property, that communications within a community have
length bounded by $O(\log\log n)$, and that the local influence of a node is within
$O(\log\log n)$ steps in the network.  The later property can be used to define some locally collective notions of networks. These observations lead to new issues of networks which will be
further discussed in Section \ref{sec:newissues}.

Secondly, we have the following:

\begin{theorem} (Community structure theorem)\label{thm:comprinciple}
For $a>1$ and $d\geq 4$, let $G$ be a network
constructed from the homophyly model. Then with probability
$1-o(1)$, the following properties hold:

\begin{enumerate}
\item [(1)] (Small community phenomenon) There are $1-o(1)$ fraction of nodes of $G$ each of which belongs to
a homochromatic set, $W$ say, such that the size of $W$ is bounded
by $O(\log^{a+1} n)$, and that the conductance of $W$, $\Phi (W)$,
is bounded by $O\left(\frac{1}{|W|^{\beta}}\right)$ for
$\beta=\frac{a-1}{4(a+1)}$.

\item [(2)] (Conductance community structure theorem) The conductance community structure ratio of $G$ is at
least $1-o(1)$, that is, $\theta (G)= 1-o(1)$.

\item [(3)] (Modularity
community structure theorem ) The modularity of $G$ is $1-o(1)$, that
is, $\sigma(G)=1-o(1)$.

\item [(4)] (Entropy community
structure theorem) The entropy community structure ratio of $G$
is $1-o(1)$, that is, $\tau(G)=1-o(1)$.

\end{enumerate}

\end{theorem}

(1) means that a set of nodes $X$ forms a natural community if the nodes in the set share the same color, that the conductance of a community $X$
is bounded by a number proportional to $|X|^{-\beta}$ for some constant $\beta$, and that communities of a network are interpretable.
(2) - (4) show that the definitions of modularity-, entropy- and conductance-
community structure are equivalent in defining community structures in networks, and that community structures are
provably definable in networks. The essence of this theorem is that community structures are definable in networks, and that communities of a network are interpretable,
giving rise to a mathematical understanding of both community structures and communities.

The fundamental and community structure principles explore some basic laws governing both the local and global structures of a network.
However, to understand the roles of community structures in networks, we need to know the properties which hold for all the communities of a network.
We will see that homophyly networks do satisfy a number of such principles.

Our third principle consists of a number of properties of degrees of the networks.
Given a node $v\in V$, we define the length of degrees of $v$ to be
the number of colors associated with all the neighbors of $v$,
written by $l(v)$. For $j\leq l(v)$, we define the $j$-th degree of
$v$ to be the $j$-th largest number of edges of the form $(v,u)$'s such that the $u$'s here
share the same color, denoted by $d_j(v)$. Define the degree of $v$,
$d(v)$, to be the number of edges incident to node $v$.

In a sharp contrast to classic graph theory, for a network
constructed from our homophyly model, $G$ say, and a vertex $v$ of
$G$, $v$ has a {\it priority of degrees}. This new feature must be
universal in real networks in the following sense: A community is an
interpretable object in a network such that nodes of the same
community share common features. In this case, a vertex $v$ may have
its own community and may link to some neighbor communities by some
priority ordering. In our model, a node $v$ more likes to contact
with nodes sharing the same color (or feature) with it, and has no
much preferences in contacting with nodes in its neighbor
communities.

For the degree priority, we have:

\begin{theorem} (Degree priority theorem)\label{thm:deprinciple} Let $G=(V,E)$ be a homophyly network. Then with probability $1-o(1)$, the
degree priority of nodes in $V$ satisfies the following properties:

\begin{enumerate}

\item [(1)] (First degree property) The first degree of $v$,
$d_1(v)$ is the number of edges from $v$ to nodes of the same color
as $v$.

\item [(2)] (Second degree property) The second degree of $v$ is
bounded by a constant, i.e., $d_2(v)\leq O(1)$

\item [(3)] (The length of degrees)
\begin{enumerate}

\item The length of degrees of $v$ is bounded by $O(\log n)$.

\item Let $N$ be the number of seed nodes in $G$. For
$r=\frac{N}{\log^cN}$ for some constant $c$.
Let $x$ be a node created after time step $r$. Then the length
of degrees of $x$ is bounded by $O(\log\log n)$.

\end{enumerate}

\item [(4)] If $v$ is a seed node, then the first degree of $v$,
$d_1(v)$ is at least $\Omega (\log^{\frac{a+1}{4}} n)$.

\end{enumerate}
\end{theorem}

\smallskip

Theorem \ref{thm:deprinciple} shows that the highest priority of a node is to link nodes of its own community, that the links of a node to nodes outside of its own community
are evenly distributed among a few communities, that for almost all nodes $x$, $x$ links to nodes of at most $O(\log\log n)$ many communities, and that for almost all seed nodes $x$,
$x$ has  at least  $\Omega (\log^{\frac{a+1}{4}} n)$ many edges linking to nodes of its own community. These properties intuitively capture the patterns of links among different communities.
Clearly, both lower and upper bounds of the length of degrees, of first and second degrees of nodes are essential to the roles of community structures of networks. For homophyly networks, we have
Theorem \ref{thm:deprinciple}. In some applications, we may need some lower bounds of the length of degrees of seeds or hubs. Anyway, the notion of degree priority provides new insight on understanding
the properties and roles of community structures of networks.

Our fourth principle determines the ways of connections from a community to outside of the community.
Let $X$ be a homochromatic set of $G$. Define the {\it width of $X$
in $G$} to be the number of nodes $x$'s such that $x\in X$ and
$l(x)>1$. We use $w^G(X)$ to denote the width of $X$ in $G$.

Then we have:

\begin{theorem}\label{thm:widthsp} ( Widths Principle): Let $G=(V,E)$ be a homophyly network. Then with probability $1-o(1)$, the following
properties hold:

\begin{enumerate}

\item [(1)] For a randomly chosen $X$, the width of $X$ in $G$ is
$w^G(X)=O(\log n)$.

Let $N$ be the number of seed nodes in $G$. For $l=N^{1-\theta}$ and
$r=\frac{N}{\log^cN}$ for some constants $\theta$ and $c$. We say
that a community is created at time step $t$, if the seed node of
the community is created at time step $t$.

\item [(2)] Let $X$ be a community created before time step $l$. Then the
width of $X$ in $G$ is at least $\Omega (\log n)$.

\item [(3)] Let $Y$ be a community created before time step $r$. Then the
width of $Y$ in $G$ is at least $\Omega (\log\log n)$

\item [(4)] Let $Z$ be a community created after time step $r$. Then the
width of $Z$ in $G$ is at most $O(\log\log n)$.

\end{enumerate}

\end{theorem}

The width of a community $X$ determines the patterns of links from
nodes in the community to nodes outside of the community.
By (4), we have that almost all communities have widths bounded by $O(\log\log n)$.
This property, together with the holographic law in the fundamental principle show that almost surely, a community has
both an internal and an external centrality. This helps us to
analyze the communications among different communities.

Our fifth principle is an inclusion and infection among the nodes of a homophyly network.
Given a node $x$ of some community $X$. We define the width of
$x$ in $G$, denoted by $w^G(x)$, is the number of communities $Y$'s such that $X\not=Y$ and such that there
is a non-seed node $y\in Y$ with which there is an edge between $x$
and $y$. Then we have:

\bigskip

\begin{theorem} \label{thm:in-in-p} ( Inclusion and infection principle): Let $G=(V,E)$ be a homophyly network. Then the following properties hold:

\begin{enumerate}
\item [(1)] (Inclusion property) For a non-seed node $x$ in $G$, the width of $x$ in $G$ is $w^G(x)=0$.

\item [(2)] (Widths of seed nodes) For every seed node $x$ in $G$, the width of $x$ is
bounded by $O(1)$.

\end{enumerate}

\end{theorem}

Intuitively speaking, non-seed nodes of a network are vulnerable against
attacks. In the cascading failure model of attacks, it is possible that a few number of attacks
may generate a global failure of the network. For this, one of the reasons is that the huge number of vulnerable nodes form
a giant connected component of the network, in which the attack of a few vulnerable nodes may infect the giant connected component of
the vulnerable nodes. (1) ensures that this is not going to happen in homophyly networks.
We interpret seed nodes as strong against attacks. Let $x$ be a seed node. If $w^G(x)>1$, then it is possible for $x$ to infect two vulnerable nodes, $y_1$ and $y_2$ say, of two different communities
$Y_1$ and $Y_2$ respectively. In this case, it is easy for $y_1$ and $y_2$ to infect the seed nodes of $Y_1$ and $Y_2$ respectively. By this way, the infections of communities intrigued by the seed node $x$ may
grow exponentially in a tree of communities. (2) ensures that for each seed node $x$ of $G$, $w^G(x)=O(1)$, which is probably larger than $1$. By this reason, we know that homophyly networks are
insecure against attacks in the cascading failure models. This suggests that to make a network $G$ secure, we have to make sure that for each hub, $x$ say, the width of $x$ in $G$ is at most $1$.
In fact, by using this principle, we have proposed a protocol of provable security of networks \cite{LPZ2013a}.

Our sixth principle is the remarkable role of seed nodes in the corresponding communities and in the whole network. We have:

\begin{theorem}\label{thm:kingp}( King node principle): Let $G=(V,E)$ be a homophyly network. Then with probability $1-o(1)$, for a community $X$ and its seed node $x_0$,
the expectation of the degree of $x_0$ is at least twice of that of the second largest degree node $x\in X$.

\end{theorem}

This principle ensures that there is a significant fraction of communities, each of which contains a king node whose degree is at least twice of that of the second largest degree node within the community.
This is a phenomenon  similar to that in a community of honey bees. It implies that
in evolutionary prisoner's dilemma games in a network, the strategies of nodes within a community could follow that of the king node,
similarly to the behaviors of a community of honey bees in nature. By using this idea, we are able to develop a theory to solve the prisoner's dilemma
in power law networks \cite{LPY2013a}.

The six principles above explore the mathematical properties of the
homophyly networks. They show that the community structures and properties of the communities do play essential roles
in fundamental issues and applications of networks.

Our model demonstrates that dynamic and scale-free
networks may have a community structure for which homophyly and
preferential attachment are the underlying mechanisms. This explains
the reason why most real networks have community structures and
simultaneously follow a power law and have a small world property.

\section{Fundamental Theorem} \label{sec:fundthm}

In this section, we prove Theorem \ref{thm:fundtheorem}.

We use $\mathcal{H}(n,a,d)$ to denote the
set of all networks of $n$ nodes constructed by the homophyly model with homophyly exponent $a$, and average number of edges $d$.

Given a network $G=(V,E)$ of the homophyly model, then every node $v\in V$ is associated with a color. The vertices $V$ is
partitioned naturally by the homochromatic sets of $V$. For an edge $e=(u,v)$, we call $e$ a {\it local edge}, if the two endpoints $u$, $v$ share the same color,
and {\it global edge}, otherwise.

A homochromatic set, $X$ say, of $V$ is expected to be a natural community of $G$. Then every community contains a seed node, which is the first
node of the community.

\begin{proof} (Proof of Theorem \ref{thm:fundtheorem})
At first, we state a Chernoff bound which will be used frequently in our proofs.

\begin{lemma} (Chernoff bound, \cite{C81}) Let $X_1,\ldots,X_n$ be independent random variables
with $\Pr[X_i=1]=p_i$ and $\Pr[X_i=0]=1-p_i$. Denote the sum by
$X=\sum\limits_{i=1}^n X_i$ with expectation $E(X)=\sum\limits_{i=1}^n p_i$. Then
we have
$$\Pr[X\leq E(X)-\lambda]\leq \exp\left(-\frac{\lambda^2}{2E(X)}\right),$$
$$\Pr[X\geq E(X)+\lambda]\leq \exp\left(-\frac{\lambda^2}{2(E(X)+\lambda/3)}\right).$$
\end{lemma}

Let $G$ be a homophyly network. We use $G_t$ to denote the graph obtained at the end of time step
$t$ of the construction of $G$, and $C_t$ to denote the set of seed nodes of $G_t$.

Let $T_1=\log^{a+1} n$.

\bigskip
For (1)(i). It suffices to show that the size of $C_t$ is bounded as
desired.

\begin{lemma} (Number of seeds lemma)\label{lem:colorsize}
With probability $1-o(1)$, for all $t\geq T_1$, $\frac{t}{2\log^a
t}\leq |C_t|\leq \frac{2t}{\log^a t}$.
\end{lemma}

By the construction of $G$, the expectation of $|C_t|$ is
$$E[|C_t|] = 2+ \sum_{i=3}^{t}\frac{1}{\log^a i}.$$
By indefinite integral
$$\int (\frac{1}{\log^a x}-\frac{a}{\log^{a+1}x}) dx=\frac{x}{\log^a x}+C,$$
we know that if $t$ is large enough, then
\begin{eqnarray*}
\sum_{i=3}^t \frac{1}{\log^a i} & \leq &
1+\int_2^t \frac{1}{\log^a x}dx\\
& \leq & \int_2^t \frac{6}{5}(\frac{1}{\log^a
x}-\frac{a}{\log^{a+1}x}) dx\\
& \leq & \frac{4t}{3\log^a t},
\end{eqnarray*}
where $\frac{6}{5}$ and $\frac{4}{3}$ are chosen arbitrarily among
the numbers larger than $1$. Similarly,
\begin{eqnarray*}
\sum_{i=3}^t \frac{1}{\log^a i} & \geq &
\int_2^t \frac{1}{\log^a x}dx\\
& \geq & \int_2^t (\frac{1}{\log^a
x}-\frac{a}{\log^{a+1}x}) dx\\
& \geq & \frac{3t}{4\log^a t}.
\end{eqnarray*}

By the Chernoff bound, with probability
$1-exp(-\Omega(\frac{t}{\log^a t}))=1-o(n^{-1})$, we have
$\frac{t}{2\log^a t}\leq |C_t|\leq \frac{2t}{\log^a t}$. By the
union bound, such an inequality holds for all $t\geq T_1$ with
probability $1-o(1)$.

(1)(i) follows from Lemma \ref{lem:colorsize} immediately.

\bigskip

Now we define an event $\mathcal{E}$:

  $|C_t|$ is bounded in the
interval $\left[\frac{t}{2\log^a t}, \frac{2t}{\log^a t}\right]$.

\smallskip

By
Lemma \ref{lem:colorsize}, $\mathcal{E}$ almost surely holds for all
$t\geq T_1$.

\bigskip
For (1)(ii). We prove the following:

\begin{lemma} (Size of community lemma)\label{lem:homosetsizeupper}
For every $T\geq T_1$, with probability $1-o(1)$, every
homochromatic set in $G_T$ has size bounded by $O(\log^{a+1}T)$.
\end{lemma}

It suffices to show that with probability $1-o(T^{-1})$, the
homochromatic set of the first color $\kappa$ has size
$O(\log^{a+1}T)$.

We define an indicator random variable $Y_t$ for the event that the
new node created at time step $t$ chooses color $\kappa$. We define
independent Bernoulli trails $\{Z_t\}$ satisfying

$$\Pr[Z_t=1]=\left(1-\frac{1}{\log^a T}\right)\frac{2\log^a
t}{t}.$$

So conditioned on the event $\mathcal{E}$ (which happens
with extremely high probability), $Y:=\sum_{t=1}^T Y_t$ is
stochastically dominated by $Z:=\sum_{t=1}^T Z_t$, which has an
expectation

$$E[Z]\leq \sum_{t=1}^T\frac{2\log^a
t}{t}\leq 2\log^{a+1} T.$$

By the Chernoff bound,
$$\Pr[Z>4\log^{a+1} T] \leq T^{-1}.$$

Therefore, with probability $1-T^{-1}$, the size of $X_\kappa$, the community of color $\kappa$, is
$Y\leq 4\log^{a+1}T$. The lemma follows.

(1)(ii) follows from Lemma \ref{lem:homosetsizeupper} by choosing
$T=n$.

\bigskip

Before proving (2) of Theorem \ref{thm:fundtheorem}, we establish both a lower and an upper bound of sizes of communities.

Set $T_2=(1-\delta_1)n$, where $\delta_1=\frac{10}{\log^{a-1}n}$.

By using a similar analysis to that in the proof of Lemma
\ref{lem:homosetsizeupper}, we have:

\begin{lemma} (Lower and upper bounds of sizes of communities)\label{lem:setsize}
With probability $1-o(1)$, both (1) and (2) below hold:
\begin{enumerate}
\item [(1)] For a homochromatic set created at a time step $\leq T_2$, it has size at least $\log
n$;
\item [(2)] For a homochromatic set created at a time step $>T_2$, it has size at most $30\log n$.
\end{enumerate}
\end{lemma}

Then we turn to the proof of the power law degree distributions in $G$,
i.e., (2) of Theorem \ref{thm:fundtheorem}.

 We prove the following
two items together, which proves (2)(i) and (2)(ii) of Theorem
\ref{thm:fundtheorem}, respectively.

\begin{enumerate}
\item [(A)] For each homochromatic set $X$, the degree distribution of induced subgraph $G_X$
follows a power law, and
\item [(B)] For each homochromatic set $X$, the degrees of nodes in $X$ follow a power law.
\end{enumerate}

(2)(iii) of Theorem \ref{thm:fundtheorem} follows immediately from
(B) by observing that the join of several power law distributions
with the same power exponent is also a power law distribution.

(2)(iv) follows from the proofs of (i) - (iii).

The idea of the proofs of (A) and (B) is to verify that the
contribution of degrees of a community from global edges is negligible, compared with those from its local
edges. This is intuitively true, since the construction of a community
basically follows the classic preferential attachment scheme, and the number of global edges created by seed nodes
is negligible. We will realize this idea gradually in the proofs below.

Let $X$ denote a homochromatic set of a fixed color, $\kappa$ say.
Let $T_0$ be the time step at which $X$ is created. Suppose that the
size of $X$ goes to infinity as $n\rightarrow\infty$. In fact, by
Lemma \ref{lem:setsize}, this holds for all the homochromatic sets
created before time $T_2$. For positive integers $s$ and $k$, define
$A_{s,k}$ to be the number of nodes of degree $k$ in $X$ when $|X|$
reaches $s$, $B_{s,k}$ to be the number of nodes of degree $k$ in
the induced subgraph of $X$ when $|X|$ reaches $s$, and $g_{s,k}$ to
be the number of global edges associated with the nodes in $X$ of
degree $k$ in the induced subgraph of $X$ when $|X|$ reaches $s$.
Obviously, $A_{s,k}=B_{s,k}+g_{s,k}$, $A_{1,d}=1$, $A_{1,k}=0$ for
all $k>d$, and $B_{1,k}=0$ for all $k$. Then we establish the
recurrence formula for the expectation of both $A_{s,k}$ and
$B_{s,k}$.

Define $T(s)$ (or $T$, for simplicity) to be the time step at which
the size of $X$ becomes to be $s$, and $s_1$ to be the number of
global edges connecting to $X$ in the case that $|X|=s$ (note that
probably at several consecutive time steps, $|X|$ keeps $s$). We
consider the time interval $(T(s-1), T(s))$. Since
$T(s)-T(s-1)=\Theta(|C_T|)=\Theta(T/\log^a T)$, the number of times
that a global edge is created and linked to a node in $X$ of degree
$k$ at some time step in the interval $(T(s-1), T(s))$ is expected
to be $\Theta(\frac{1}{\log^a T}\cdot\frac{dk\cdot
A_{s,k}}{2dT}\cdot\frac{T}{\log^a T})=\Theta(\frac{k\cdot
A_{s,k}}{\log^{2a}T})$. Denote $\Theta(\log^{2a}T)$ by $s_2$.

For $s>1$ and $k>d$, we have
$$E(A_{s,k})=A_{s-1,k}\left(1-\frac{kd}{2d(s-1)+s_1}-\frac{k}{s_2}\right)
+A_{s-1,k-1}\cdot\left(\frac{(k-1)d}{2d(s-1)+s_1}+\frac{k-1}{s_2}\right)+O\left(\frac{1}{s^2}\right),$$
where the error terms caused by the case that more than one edge
joins to a single node are absorbed in the $O(1/s^2)$ term. Taking
expectations on both sides, we have
\begin{eqnarray} \label{eqn:A_krecurrence}
E(A_{s,k}) &=&
E(A_{s-1,k})\left(1-\left(\frac{1}{2(s-1)+s_1/d}-\frac{1}{s_2}\right)k\right)\nonumber \\
&&+E(A_{s-1,k-1})\left(\frac{1}{2(s-1)+s_1/d}+\frac{1}{s_2}\right)(k-1)+O\left(\frac{1}{s^2}\right).
\end{eqnarray}
When $k=d$,
\begin{equation} \label{eqn:A_drecurrence}
E(A_{s,d})=E(A_{s-1,d})\left(1-\left(\frac{1}{2(s-1)+s_1/d}-\frac{1}{s_2}\right)d\right)+1+O\left(\frac{1}{s^2}\right).
\end{equation}

Similarly, for $s>1$ and $k>d$,
$$E(B_{s,k})=B_{s-1,k}-\frac{d\cdot(kB_{s-1,k}+g_{s-1,k})}{2d(s-1)+s_1}+\frac{d\cdot((k-1)B_{s-1,k-1}+g_{s-1,k-1})}{2d(s-1)+s_1}+O(\frac{1}{s^2}).$$
Taking expectations on both sides, we have
\begin{eqnarray} \label{eqn:B_krecurrence}
E(B_{s,k}) &=& E(B_{s-1,k})\left( 1-\frac{kd}{2d(s-1)+s_1}
\right)+E(B_{s-1,k-1})\cdot
\frac{(k-1)d}{2d(s-1)+s_1}\nonumber \\
&&+\frac{E(g_{s-1,k-1}-g_{s-1,k})}{2d(s-1)+s_1}+O(\frac{1}{s^2}).
\end{eqnarray}
When $k=d$,
\begin{eqnarray} \label{eqn:B_drecurrence}
E(B_{s,d}) &=&
B_{s-1,d}-\frac{d\cdot(dB_{s-1,d}+g_{s-1,d})}{2d(s-1)+s_1}+1+O(\frac{1}{s^2})\nonumber \\
&=& B_{s-1,d}\left( 1-\frac{d}{2(s-1)+s_1/d} \right)+\left(
1-\frac{g_{s-1,d}}{2d(s-1)+s_1} \right),
\end{eqnarray}
and
$$E(B_{s,d})=E(B_{s-1,d})\left( 1-\frac{d}{2(s-1)+s_1/d} \right)+\left(
1-\frac{E(g_{s-1,d})}{2d(s-1)+s_1} \right).$$

To solve these recurrences, we introduce the following lemma that is
used in the canonical proof of the preferential attachment model.

\begin{lemma}
\label{lem:recurrence} (\cite{CL06}, Lemma 3.1) Suppose that a
sequence $\{a_s\}$ satisfies the recurrence relation
$$a_{s+1}=(1-\frac{b_s}{s+s_1})a_s+c_s~~{\it for}~~s\geq s_0,$$
where the sequences $\{b_s\},\{c_s\}$ satisfy
$\lim_{s\rightarrow\infty}b_s=b>0$ and
$\lim_{s\rightarrow\infty}c_s=c$ respectively. Then the limit
of $\frac{a_s}{s}$ exists and
$$\lim_{s\rightarrow\infty}\frac{a_s}{s}=\frac{c}{1+b}.$$
\end{lemma}

For the recurrence of $E(A_{s,k})$, we have to deal with $s_1$ and
$s_2$. Note that $s_2=\Theta(\log^{2a}T)=\omega(s)$. For $s_1$, we
give an upper bound for the expected volume of $X$ at time $T$,
denoted by $V_T$, as follows.
\begin{eqnarray*}
E(V_T) &=& \sum\limits_{i=2}^T \left[ \left( 1-\frac{1}{\log^a i}
\right)\cdot\frac{2d}{|C_i|}+\frac{1}{\log^a i}\cdot\frac{d
V_{i-1}}{2di} \right]\\
&=& O\left(\sum\limits_{i=2}^T \frac{2d}{|C_i|}\right) =
O\left(\sum\limits_{i=2}^T \frac{4d\log^a i}{i}\right) =
O(\log^{a+1} T).
\end{eqnarray*}

In fact, by using an analysis of martingale, we are able to show that, almost surely, almost all
homochromatic sets have size at most $O(\log^{a+1} T)$. So it is easy
to observe that $\frac{s_1}{t}=O\left( \frac{1}{\log^a
T}\cdot\frac{V_T}{2dT}/ \frac{\log^a T}{T} \right)=O\left(
\frac{1}{\log^{a-1} T} \right)$ goes to zero as $n$, and in turn
$s$, approach to infinity. For the recurrence of $E(B_{s,k})$, we
show that as $s$ goes to infinity, both
$\frac{E(g_{s-1,k-1}-g_{s-1,k})}{2d(s-1)+s_1}$ and
$\frac{E(g_{s-1,d})}{2d(s-1)+s_1}$ approach to $0$. Define
$g_s=\sum_i g_{s,i}$ to be the total number of global edges
associated to $X$ when $|X|$ reaches $s$. We show that
$E(\frac{g_s}{s})\rightarrow 0$ as $s\rightarrow \infty$.

Note that $X$ is created at time $T_0$.
\begin{eqnarray*}
E(g_s) = O\left( \sum\limits_{i=T_0}^{T(s)}\frac{1}{\log^a
i}\cdot\frac{dV_i}{2di} \right) = O\left(\sum\limits_{i=T_0}^{T(s)}
\frac{\log i}{2i}\right) = O(\log^2 T(s)-\log^2 T_0).
\end{eqnarray*}
We consider the size of $X$ at some time step $t>T_0$.
\begin{eqnarray*}
E(|X|) &=& \sum\limits_{i=T_0}^{t} \left( 1-\frac{1}{\log^a i}
\right)\frac{1}{|C_i|} = \Omega\left(
\sum\limits_{i=T_0}^{t}\frac{\log^a i}{2i} \right)\\
&=& \Omega\left( \int_{T_0}^{t}\frac{\log^a x}{2x}dx \right) =
\Omega(\log^{a+1}t-\log^{a+1}T_0).
\end{eqnarray*}
Thus at time $T(s)$, by the Chernoff bound, with probability
$1-o(1)$, $s=\Omega(\log^{a+1}T(s)-\log^{a+1}T_0)$. Therefore,
$E(g_s)=o(s)$, that is, $E(\frac{g_s}{s})\rightarrow 0$ as
$s\rightarrow \infty$.

Then we turn to solve the recurrences of $E(A_{s,k})$ and
$E(B_{s,k})$. Now the terms $s_1/d$ and $\frac{1}{s_2}$ in
equalities (\ref{eqn:A_krecurrence}) and (\ref{eqn:A_drecurrence})
are comparatively negligible, and so do the terms
$\frac{E(g_{s-1,k-1}-g_{s-1,k})}{2d(s-1)+s_1}$ and
$\frac{E(g_{s-1,d})}{2d(s-1)+s_1}$ in equalities
(\ref{eqn:B_krecurrence}) and (\ref{eqn:B_drecurrence}). By Lemma
\ref{lem:recurrence}, $\frac{E(A_{s,k})}{s}$ and
$\frac{E(B_{s,k})}{s}$ must have the same limit as $t$ goes to
infinity. Thus we will only give the proof of the power law
distribution for $E(A_{s,k})$, which also holds for $E(B_{s,k})$.

Denote by $S_k=\lim_{t\rightarrow\infty}\frac{E(A_{s,k})}{s}$ for
$k\geq d$. In the case of $k=d$, we apply Lemma \ref{lem:recurrence}
with $b_s=d/2$, $c_s=1+O(1/s)$, $s_1=-1$, and get
$$S_d=\lim_{s\rightarrow\infty}\frac{E(A_{s,d})}{t}=\frac{1}{1+\frac{d}{2}}=\frac{2}{2+d}.$$
For $k>d$, assume that we already have
$S_{k-1}=\lim_{t\rightarrow\infty}\frac{E(A_{s,k-1})}{t}$. Applying
Lemma \ref{lem:recurrence} again with $b_s=k/2$,
$c_s=\frac{E(A_{s-1,k-1})}{s-1}\cdot\frac{k-1}{2}$, $s_1=-1$, we get
$$S_k=\lim_{t\rightarrow\infty}\frac{E(A_{s,k})}{s}=\frac{S_{k-1}\cdot\frac{k-1}{2}}{1+\frac{k}{2}}=S_{k-1}\cdot\frac{k-1}{k+2}.$$
Thus recurrently, we have
\begin{equation} \label{S_k expression}
S_k=S_d\cdot\frac{(d+2)!(k-1)!}{(d-1)!(k+2)!}=\frac{2d(d+1)}{k(k+1)(k+2)}.
\end{equation}
This implies
$$|E(A_{s,k})-S_k \cdot s|=o(s),$$
and thus
$$E(A_{s,k})=(1+o(1))k^{-3}s.$$
Since $s=\omega(1)$ goes to infinity as $n\rightarrow\infty$,
$E(A_{s,k})\propto k^{-3}$. For the same reason, $E(B_{s,k})\propto
k^{-3}$. This proves (A) and (B), and also completes the proof of
(2)(i) and (2)(ii).

\bigskip
For (2)(iii), a key observation is that the union of several power
law distributions is also a power law distribution if the power exponents are
equal. We will give the same explicit expression of the expectation
of the number of degree $k$ nodes by combining those for the
homochromatic sets, leading to a similar power law distribution.

To prove the power law degree distribution of the whole graph, we
take the union of distributions of all homochromatic sets. We will
show that with overwhelming probability, almost all nodes belong to
some large homochromatic sets so that the role of small
homochromatic sets is unimportant.

Suppose that $G_n$ has $m$ homochromatic sets of size at least $\log
n$. For $i=1,\ldots,m$, let $M_i$ be the size of the $i$-th
homochromatic set and $N_{s,k}^{(i)}$ denote the number of nodes of
degree $k$ when the $i$-th set has size $s$. For each $i$, we have
$$\lim_{n\rightarrow\infty}\frac{E(N_{M_i,k}^{(i)})}{M_i}=S_k.$$
Hence,
$$\lim_{n\rightarrow\infty}\frac{E(\sum_{i=1}^m
N_{M_i,k}^{(i)})}{\sum_{i=1}^m M_i}=S_k.$$

Let $M_0$ denote the size of the union of all other homochromatic
sets of size less than $\log n$, and $N_{s,k}^{(0)}$ denote the
number of nodes of degree $k$ in this union when it has size $s$. By
Lemma \ref{lem:setsize}, with probability $1-o(1)$, all these sets
are created after time $T_2$, and thus $M_0\leq n-T_2=
\frac{10n}{\log^{a-1}n}=o(n)$.

Define $N_{t,k}$ to be the number of nodes of degree $k$ in $G_t$,
that is, the graph obtained after time step $t$. Then we have
$$\lim_{n\rightarrow\infty}\frac{E(N_{n,k})}{n}=\lim_{n\rightarrow\infty}\frac{E(\sum_{i=0}^m N_{M_i,k}^{(i)})}{\sum_{i=0}^m M_i}.$$
For $M_0$, we have that
$$\lim_{n\rightarrow\infty} \frac{M_0}{\sum_{i=1}^m M_i}=\lim_{n\rightarrow\infty} \frac{M_0}{n-M_0}=0$$
and
$$\lim_{n\rightarrow\infty} \frac{E(N_{M_0,k}^{(0)})}{n} \leq \lim_{n\rightarrow\infty} \frac{M_0}{n}=0$$
hold with probability $1-o(1)$. So
$$\lim_{n\rightarrow\infty} \frac{E(N_{n,k})}{n}=\lim_{n\rightarrow\infty}\frac{E(\sum_{i=1}^m N_{M_i,k}^{(i)})}{\sum_{i=1}^m M_i}=S_k.$$
This implies
$$|E(N_{n,k})-S_k\cdot n|=o(n),$$
and thus,
$$E(N_{n,k})=(1+o(1))k^{-3}n,$$
and $E(N_{n,k})\propto k^{-3}$. (2)(iii) follows.

(2)(iv) is clear from the proofs of (2)(i) - (2)(iii).

This completes the proof of Theorem \ref{thm:fundtheorem}(2).

\bigskip
Then we turn to the proof of the third part of Theorem \ref{thm:fundtheorem}.

For (3)(i), we will use the well-known result on the diameter of a
graph from the PA model to bound the diameter of each homochromatic
set. Bollob{\'a}s and Riordan \cite{BR04} have shown that a randomly
constructed graph of size $n$ from the PA model has a diameter
$O(\log n)$ with probability $1-O(\frac{1}{\log^2 n})$. By Lemma
\ref{lem:homosetsizeupper}, we know that the sizes of all
homochromatic sets are bounded by $O(\log^{a+1} n)$. Thus the
induced subgraph of a homochromatic set has diameter $O(\log\log
n)$. (3)(i) follows.

\bigskip
For (3)(ii), to consider the average node to node distance of the
whole graph $G$, we first clarify the hierarchical structure of $G$
as follows. The first level of $G$ is obtained by shrinking the
nodes of the same color in $G$ to a single node while maintaining
the global edges. Denote the first-level graph by $G'$. The second
level of $G$ is the graph obtained from $G$ by simply deleting all the global edges
from $G$, which consists of the isolated homochromatic sets.

We define a path $P_{u,v}$ connecting two nodes $u,v$ as follows. If
$u$ and $v$ share the same color, then $P_{u,v}$ is the shortest path
between $u$ and $v$ in the corresponding homochromatic set.
Otherwise, choose the shortest path from $u$ to the seed node $s_u$
in the same homochromatic set and pick among the $d$ global edges
born with $s_u$ the one which connects to the earliest created node,
say $u'$. These two parts compose a path $P_{u,u'}$ from $u$ to
$u'$. Do the same to $v$ and also find a path $P_{v,v'}$.
Recursively, we define the path $P_{u',v'}$, and $P_{u,v}$ consists
of $P_{u,u'}$, $P_{u',v'}$ and $P_{v,v'}$.

Note that $P_{u,v}$ consists of paths from the two levels of $G$
alternately, that is, $P_{u,v}$ consists of blocks of local edges
and global edges alternately. Next, we consider the paths in the two
levels and show that the average node to node path $P_{u,v}$ has
length at most $O(\log^2 n)$ with high probability.

To estimate the number of edges in $P_{u,v}$ from $G'$, i.e., the
first-level graph of $G$, we recall a known conclusion on random
recursive trees. A random recursive tree is constructed by stages.
At each stage, a new vertex is created and linked to an earlier node
uniformly and randomly. In this case, we call it a uniform recursive
tree \cite{MS95}. We use a result of Pittel in \cite{P94}, saying
that the height of a uniform recursive tree of size $n$ is $O(\log
n)$ with high probability.

\begin{lemma} (Recursive tree lemma) \label{lem:resursivetrees}
(\cite{P94}) With probability $1-o(1)$, the height of a uniform
recursive tree of size $n$ is asymptotic to $e\log n$, where $e$ is
the natural logarithm.
\end{lemma}

Consider $G'$ as a union of $d$ recursive trees. Note that the
earlier created homochromatic sets in $G$ have larger expected
volumes than the later created ones. So with higher probability than
the uniform recursive tree, the height of a recursive tree in $G'$ is
asymptotic to $e\log |C_n|$, where $|C_n|$ is the number of colors
in $G$ and is also the number of nodes in $G'$. This means that with
probability $1-o(1)$, the number of global edges in $P_{u,v}$ is at
most $2e\log |C_n|=O(\log n)$.

To estimate the diameters of the homochromatic sets in the
second-level graph of $G$, we adjust the parameters in the proof of
the diameter of the PA model in \cite{BR04} to get a weaker bound on
diameters, but a tighter bound on probability. In so doing, we have
the following lemma.

\begin{lemma} (Diameter of PA networks) \label{lem:padiam}
For any constant $a'>2$, there is a constant $K$ such that with
probability $1-\frac{1}{n^{a'+1}}$, a randomly constructed graph $G$
from the PA model $\mathcal{P}(n,d)$ has a diameter $Kn^{1/(a'+1)}$.
\end{lemma}

The proof for this is a standard argument as that in the proof of the small diameter
property of networks of the PA model.

Choose $a'$ in Lemma~\ref{lem:padiam} to be the homophyly exponent
$a$, and then we have a corresponding $K$ from Lemma
\ref{lem:padiam}. Given a homochromatic set $S$, we say that $S$ is
{\it bad}, if the diameter of $S$ is larger than $K|S|^{1/(a+1)}$.
We define an indicator $X_S$ of the event that $S$ is bad. Since
$\log n\leq|S|=O(\log^{a+1}n)$, by Lemma \ref{lem:padiam}, for a
randomly chosen $S$,
$$\Pr[X_S=1]\leq \frac{1}{\log^{a+1}n}.$$
By Lemma \ref{lem:colorsize}, the expected number of bad sets is at
most $\frac{2n}{\log^a
n}\cdot\frac{1}{\log^{a+1}n}=\frac{2n}{\log^{2a+1}n}$. By the
Chernoff bound, with probability $1-O(n^{-2})$, the number of bad
sets is at most $\frac{3n}{\log^{2a+1}n}$. Thus the total number of
nodes belonging to some bad set is $O\left(\frac{n}{\log^a
n}\right)$. On the other hand, for any large set $S$ that is not
bad, its diameter is at most $K|S|^{1/(a+1)}=O(\log n)$.

To estimate the average node-to-node distance of $G$, we consider
the length of $P_{u,v}$ for uniformly and randomly chosen $u$ and
$v$. If neither $u$ nor $v$ is in a bad homochromatic set, then the
length of $P_{u,v}$ is $O(\log n)$. Otherwise, its length is
$O(\log^{a+1}n \cdot \log n)=O(\log^{a+2}n)$. Thus, the average node
to node distance in $G$ is bounded by
$$O(\frac{\frac{n^2}{\log^a
n}\cdot\log^{a+2}n+n^2\cdot\log n}{n^2})=O(\log^2 n).$$ (3)(ii)
follows.

This completes the proof of Theorem \ref{thm:fundtheorem}.
\end{proof}

\section{Community Structure Principle} \label{sec:ccs}

In this section, we prove Theorem \ref{thm:comprinciple}.

We
will show that the homochromatic sets appearing not too early and
too late are good communities with high probability. Then the
theorem follows from the fact that the total size of the remaining
part of nodes takes up only $o(1)$ fraction of nodes in $G$.

Set $T_3=\frac{n}{\log^{a+2} n},
T_4=\left(1-\frac{1}{\log^{(a-1)/2}n}\right)n$.

\begin{proof} (Prof of Theorem \ref{thm:comprinciple})

For (1). We focus on the homochromatic sets created in time interval
$[T_3,T_4]$. Given a
homochromatic set $S$, we use $t_S$ to denote the time at which $S$
is created. Suppose that $S$ is a homochromatic set with
$t_S\in[T_3,T_4]$, and $s$ is the seed node of $S$. For $t\geq t_S$,
define $S[t]$ to be the snapshot of $S$ at time step $t$, and
$\partial(S)[t]$ to be the set of edges from $S[t]$ to
$\overline{S[t]}$, the complement of $S[t]$. In our proof, we first
make an estimation on the total degrees of nodes in $S[t]$ at any
given time $t>t_S$, and then show that the global edges connecting
to $S$ is not too many.

For each $t\geq t_S$, we use $D(S)[t]$ to denote the total degree of
nodes in $S[t]$ at the end of time step $t$. We have the following
lemma.

\begin{lemma} (Degree of communities lemma) \label{lem:homodegree}
For any homochromatic set $S$ created at time $t_S\geq T_3$,
$D(S)[n]=O(\log^{a+1}n)$ holds with probability $1-o(1)$.
\end{lemma}

We only have to show that for any $t\geq T_3$, if $S$ is a
homochromatic set created at time step $t$, then
$D_n(S)[n]=O(\log^{a+1}n)$ holds with probability $1-o(n^{-1})$. We
assume the worst case that $S$ is created at time step $t_S=T_3$.
The recurrence on $D(S)[t]$ can be written as
\begin{eqnarray*}
E[D(S)[t]\ |\ D(S)[t-1]] &=& D(S)[t-1]+\frac{1}{\log^a t} \cdot
\frac{D(S)[t-1]}{2d(t-1)}\cdot d \\
&& +\left(1-\frac{1}{\log^a t}\right)\cdot\frac{2d}{|C_{t-1}|}.
\end{eqnarray*}

We suppose again the event $\mathcal{E}$ that for all $t\geq
T_1=\log^{a+1}n$, $\frac{t}{2\log^a t}\leq |C_t|\leq
\frac{2t}{\log^a t}$, which almost surely happens by Lemma
\ref{lem:colorsize}. It holds also for $t\geq T_3$. On this
condition,
\begin{eqnarray}
E[D(S)[t]\ |\ D(S)[t-1],\mathcal{E}]&\leq& D(S)[t-1]
\left[1+\frac{1}{2(t-1)\log^a t}\right] + \frac{4d\log^a t}{t}.
\label{eqn:degree}
\end{eqnarray}

To deal with this recurrence, we use the submartingale concentration
inequality (see \cite{CL06}, Chapter 2, for information on
martingales) to show that $D(S)[t]$ is small with high probability.

Since
\begin{eqnarray*}
&&10d\log^{a+1}(t+1)-10d\left(1+\frac{1}{2(t-1)\log^a t}\right)\cdot\log^{a+1} t\\
&\geq& 10d\log^a t\left(\log\frac{t+1}{t}\right)-\frac{10d\log t}{2(t-1)}\\
&\geq& \frac{10d\log^a t}{t+1}-\frac{10d\log^a t}{2(t-1)}\\
&\geq& \frac{4d\log^a t}{t},
\end{eqnarray*}

applying it to Inequality (\ref{eqn:degree}), we have
\begin{eqnarray*}
&&E[D(S)[t]\ |\ D(S)[t-1],\mathcal{E}]-10d\log^{a+1} (t+1)\\
&\leq& \left(1+\frac{1}{2(t-1)\log^a t}\right) \cdot
(D(S)[t-1]-10d\log^{a+1} t).
\end{eqnarray*}

For $t\geq T_3$, define $\theta_t=\Pi_{i=T_3+1}^{t}
\left(1+\frac{1}{2(i-1)\log^a i}\right)$ and
$X[t]=\frac{D(S)[t]-10d\log^{a+1} (t+1)}{\theta_t}$. Then
$$E[X[t]\ |\ X[t-1],\mathcal{E}]\leq X[t-1].$$

Note that
$$X[t]-E[X[t]\ |\ X[t-1],\mathcal{E}] = \frac{D(S)[t]-E[D(S)[t]\ |\ D(S)[t-1],E]}{\theta_t}\leq 2d.$$
Since
$$D(S)[t]-D(S)[t-1]\leq 2d,$$
we have
\begin{eqnarray*}
{\rm Var}[X[t]\ |\ X[t-1],\mathcal{E}] &=&
E[(X[t]-E(X[t]|X[t-1],\mathcal{E}))^2]\\
&=& \frac{1}{\theta_t^2}
E[(D(S)[t]-E(D(S)[t]\ |\ D(S)[t-1],\mathcal{E}))^2]\\
&\leq& \frac{1}{\theta_t^2}
E[(D(S)[t]-D(S)[t-1])^2|D(S)[t-1],\mathcal{E}]\\
&\leq& \frac{2d}{\theta_t^2}
E[D(S)[t]-D(S)[t-1]\ |\ D(S)[t-1],\mathcal{E}]\\
&\leq& \frac{2d}{\theta_t^2} \left[ \frac{4d\log^a
t}{t}+\frac{D(S)[t-1]}{2(t-1)\log^a t} \right]\\
&=& \frac{8d^2\log^a t}{t\theta_t^2}+\frac{d}{(t-1)\theta_t\log^a
t}\cdot\frac{D(S)[t-1]}{\theta_t}\\
&\leq& \frac{8d^2\log^a t}{t\theta_t^2}+\frac{10d^2\log^{a+1}
t}{(t-1)\theta_t^2\log^a t}+\frac{d X[t-1]}{(t-1)\theta_t\log^a
t}\\
&\leq& \frac{10d^2\log^a t}{t\theta_t^2}+\frac{d
X[t-1]}{(t-1)\theta_t\log^a t}.
\end{eqnarray*}
Note that $\theta_t$ can be bounded as
\begin{eqnarray*}
\theta_t&\sim& e^{\sum_{i=T_3+1}^t\frac{1}{2(i-1)\log^a i}} \in
[(\frac{t}{T_3})^{\frac{1}{2\log^a n}},
(\frac{t}{T_3})^{\frac{1}{2\log^a T_3}}].
\end{eqnarray*}
Then
\begin{eqnarray*}
\sum_{i=T_3+1}^t\frac{10d^2\log^a i}{i\theta_i^2}\leq 10d^2\log^a n
\int_{T_3}^t\frac{1}{x}\cdot\left(\frac{T_3}{x}\right)^{\frac{1}{\log^a
n}}dx \leq 10d^2\log^a n \cdot \log n=10d^2\log^{a+1} n,
\end{eqnarray*}
and
\begin{eqnarray*}
\sum_{i=T_3+1}^t\frac{1}{(i-1)\theta_i\log^a i}\leq \frac{2}{\log^a
T_3}\int_{T_3}^t\frac{T_3^{\frac{1}{2\log n}}}{x\cdot
x^{\frac{1}{2\log n}}}dx \leq \frac{2\log n}{\log^a T_3}.
\end{eqnarray*}

Here we can safely assume that $X[t]$ is non-negative, which means
that $D(S)[t]\geq 10\log^{a+1}(t+1)$, because otherwise, the
conclusion follows immediately. Let $\lambda=20\log^{a+1}n$. By the
submartingale inequality (\cite{CL06}, Theorem 2.40),
\begin{eqnarray*}
&&\Pr[X[t]=\omega(\log^{a+1}n)]\leq\Pr[X[t]\geq X[T_3]+\lambda]\\
&\leq& \exp(-\frac{\lambda^2}{2(10d^2\log^{a+1} n + (2\log n/\log^a
T_3)\lambda + d\lambda/3)})+O(n^{-2})=O(n^{-2}).
\end{eqnarray*}
This implies that $D(S)[n]=O(\log^{a+1}n)$ holds with probability
$1-O(n^{-2})$.

Suppose that $T_3\leq t_S<T_4$. We consider the edges from seed
nodes created after time step $t_S$ to nodes in $S$. By a similar
proof to that in Lemma \ref{lem:setsize} (1), we are able to show
that, with probability $1-o(1)$, $S=S[n]$ has a size
$\Omega(\log^{\frac{a+1}{2}} n)$, and so a volume
$\Omega(\log^{\frac{a+1}{2}} n)$. We suppose the event, denoted by
$\mathcal{F}$, that for any $t\geq T_S$, $D(S)[t]=O(\log^{a+1}n)$,
which holds with probability $1-o(1)$ by Lemma \ref{lem:homodegree}.
For each $t\geq T_S$, we define a random variable $X_t$ to be the
number of global edges that connect to $S$ at time $t$. We have
$$E[X_t|\mathcal{F}]=d \cdot \frac{1}{\log^a t} \cdot \frac{D(S)[t-1]}{2d(t-1)}\leq \frac{\log^{1+\epsilon} n}{2(t-1)},$$
for arbitrarily small positive $\epsilon$. Then
$$E[\sum_{t=t_S}^n X_t|\mathcal{F}]\leq \log^{1+\epsilon}n\sum_{t=t_S}^n\frac{1}{2(t-1)}\leq a(\log^{1+\epsilon}n)(\log\log n).$$
By the Chernoff bound,
$$\Pr[\sum_{t=t_S}^n X_t\geq 2a(\log^{1+\epsilon}n)(\log\log n)]\leq n^{-2}.$$
That is, with probability at least $1-n^{-2}$, the total number of
global edges joining $S$ is upper bounded by
$2a(\log^{1+\epsilon}n)(\log\log n)$.

Let $0<\epsilon<\frac{a-1}{4}$. Then, with probability $1-o(1)$, for
each such $S$ (satisfying $t_S\in[T_3,T_4]$), the conductance of $S$
is
$$\Phi (S)=O\left(\frac{2a(\log^{1+\epsilon}n)(\log\log n)+\log n}{\log^{(a+1)/2} n}\right)
\leq O\left(\log^{-\frac{a-1}{4}}n\right)\leq
O\left(|S|^{-\frac{a-1}{4(a+1)}}\right).$$

On the other hand, the total number of nodes belonging to the
homochromatic sets which appear before time $T_3$ or after time
$T_4$ is at most
$O(\log^{a+1}n)\cdot\frac{n}{\log^{a+2}n}+\frac{n}{\log^{(a-1)/2}
n}=o(n)$ for any constant $a>1$. Therefore, $1-o(1)$ fraction of
nodes of $G$ belongs to a subset $W$ of nodes, which has a size
bounded by $O(\log^{a+1} n)$ and a conductance bounded by
$O\left(|W|^{-\frac{a-1}{4(a+1)}}\right)$. (1) follows.

For (2).  We only
have to show that in the case of a specific $\mathcal{X}$ in $G$,
$\theta^\mathcal{X}(G)=1-o(1)$.

We define $\mathcal{X}$ by colors such that each homochromatic set
created before time $T_4$ is a module in $\mathcal{X}$. Note that
each module is connected, and by Lemma \ref{lem:homosetsizeupper}
and \ref{lem:setsize}, with probability $1-o(1)$, its size is
between $\log n$ and $\sqrt{n}$. So each module is a possible
community.

If a node $x$ is in a homochromatic set $S$ with $t_S\in[T_3,T_4]$,
$a^\mathcal{X}(x)=1-\Phi(S)=1-O\left(\log^{-\frac{a-1}{4}}n\right)$.
Otherwise, we assume the worst case that $a^\mathcal{X}(x)=0$. Since
the number of such nodes is at most $o(n)$, we have
$$\theta^\mathcal{X}(G) \geq
\frac{n-o(n)}{n}\left[1-O\left(\log^{-\frac{a-1}{4}}n\right)\right]=1-o(1).$$

(2) follows.

For (3). We define the partition $\mathcal{P}$ as follows. Each homochromatic
set $S$ with $t_S\in[T_3,T_4]$ is a module in $\mathcal{P}$, and the
union of the rest homochromatic sets, that is those created before
$T_3$ or after $T_4$, forms a module in $\mathcal{P}$.

Note that
\begin{eqnarray} \label{eqn:modularity_G}
\sigma^{\mathcal{P}}(G) = \sum\limits_{l=1}^L \left[
\frac{k_l}{m}-\left(\frac{V_l}{2m}\right)^2 \right] =
\frac{1}{m}\sum\limits_{l=1}^L k_l - \sum\limits_{l=1}^L
\left(\frac{V_l}{2m}\right)^2,
\end{eqnarray}
where $\sum_{l=1}^L
k_l$ is at least the number of local edges in $G$. Since the number
of global edges is exactly $d\cdot|C_n|$, which by Lemma
\ref{lem:colorsize} is at most $2dn/\log^{a+1}n$ with probability
$1-o(1)$, the number of local edges is $m-(2dn/\log^{a+1}n)$. Since
$m=dn$, we have
$$\frac{1}{m}\sum\limits_{l=1}^L k_l \geq 1-(2/\log^{a+1}n).$$

Next we bound $V_l$ for each module $l$. First we consider the
homochromatic sets appearing in time interval $[T_3,T_4]$. By Lemma
\ref{lem:homodegree}, with probability $1-o(1)$, the volume of every
such $l$ is bounded by $O(\log^{a+1}n)$. So the contribution of
these modules to the term $\sum_{l=1}^L
\left(\frac{V_l}{2m}\right)^2$ in Equation (\ref{eqn:modularity_G})
is $o(1)$.

Then we consider the module which is the union of the homochromatic
sets appearing before $T_3$ or after $T_4$. Since
$T_4=\left(1-\frac{1}{\log^{(a-1)/2}n}\right)n$, the total volume of
the homochromatic sets appearing after $T_4$ is at most
$\frac{2dn}{\log^{(a-1)/2}n}$. For those appearing before $T_3$,
since $T_3=\frac{n}{\log^{a+2} n}$, the total volume of them cannot
exceed $\frac{2dn}{\log^{a+2} n}O(\log^{a+1}n)=O(n/\log n)$ plus the
number of all global edges. Since the latter is at most
$2dn/\log^{a+1}n$ with probability $1-o(1)$, the volume of this part
is at most $O(n/\log n)$. So the contribution of this module to the
term $\sum_{l=1}^L \left(\frac{V_l}{2m}\right)^2$ in Equation
(\ref{eqn:modularity_G}) is also $o(1)$.

Combining these two parts, the term $\sum_{l=1}^L
\left(\frac{V_l}{2m}\right)^2$ in Equation (\ref{eqn:modularity_G})
is $o(1)$. Thus $\sigma^{\mathcal{P}}(G)=1-o(1)$.  (3) follows.

For (4). We define a partition $\mathcal{P}$ as follows: Each homochromatic
set in $G$ is a module in $\mathcal{P}$. We will calculate $L^U(G)$
and $L^{\mathcal{P}}(G)$, respectively. We will use the power law degree
distribution of $G$, and also of each module.

By Theorem \ref{thm:fundtheorem}, (2)(iii) and (2)(ii), the degrees of
nodes in $G$ follows a power law distribution with power exponent $\beta=3$,
and this holds in each homochromatic set, that is, in each module in
$\mathcal{P}$. Let
$$A=\sum\limits_{k=d}^{d_\mathrm{max}} k^{-3},$$
where $d_\mathrm{max}$ is the maximum degree of nodes in $G$. So the
number of nodes of degree $i$ in $G$ is (roughly) $n\cdot i^{-3}/A$.

Note that
$$A=\sum\limits_{k=d}^{d_\mathrm{max}} k^{-3} \leq \int_{d-1}^\infty \frac{1}{x^3}dx =
\frac{1}{2(d-1)^2}.$$

So the number of nodes of degree $i$ in
$G$ is at least $\frac{2(d-1)^2 n}{i^3}$. Therefore,

\begin{eqnarray*}
L^U(G) &\geq& -\sum\limits_{i=d}^{d_\mathrm{max}}
\left(\frac{i}{2m}\cdot\log_2\frac{i}{2m}\right)\cdot\frac{2(d-1)^2
n}{i^3}\\
&=& \frac{(d-1)^2n\log_2 e}{m}
\sum\limits_{i=d}^{d_\mathrm{max}}\frac{1}{i^2}\cdot\log\frac{2m}{i}\\
&\geq& \frac{(d-1)^2n\log_2 e}{m}
\int_d^{d_\mathrm{max}}\frac{1}{x^2}\cdot\log\frac{2m}{x}dx\\
&=& \frac{(d-1)^2n\log_2 e}{2m^2}
\int_{\frac{2m}{d_\mathrm{max}}}^{\frac{2m}{d}}\log y dy\\
&=& \frac{(d-1)^2n\log_2 e}{2m^2} \left[ \left(
\frac{2m}{d}\cdot\log\frac{2m}{d}-\frac{2m}{d} \right) - \left(
\frac{2m}{d_\mathrm{max}}\cdot\log\frac{2m}{d_\mathrm{max}}-\frac{2m}{d_\mathrm{max}}
\right) \right].\\
\end{eqnarray*}
Since $d_\mathrm{max}$ goes to infinity as $m\rightarrow\infty$,
$\frac{2m}{d}\gg\frac{2m}{d_\mathrm{max}}$ for large enough $m$.
Note that $m=dn$. So we have
\begin{eqnarray*}
L^U(G) &\geq& \frac{(d-1)^2n}{2m^2}\cdot\frac{2m}{d}\cdot
\log\frac{2m}{d}\\
&\geq& \left(\frac{d-1}{d}\right)^2 \log n.
\end{eqnarray*}

Then we give an upper bound for $L^{\mathcal{P}}(G)$. For each
homochromatic set $j$, let $L_j=-\sum\limits_{i=1}^{n_j}
\frac{d_i^{(j)}}{V_j} \cdot \log_2\frac{d_i^{(j)}}{V_j}$ and
$L_\mathcal{P}=-\sum\limits_{j=1}^L \frac{V_j}{2m} \cdot
\log_2\frac{V_j}{2m}$. By the definition of $L^{\mathcal{P}}(G)$,

$$L^{\mathcal{P}}(G)=\sum\limits_{j=1}^L \frac{V_j}{2m} L_j + \frac{m_g}{m} L_\mathcal{P}.$$

To bound $L_j$'s and $L_\mathcal{P}$, we note that, by information
theoretical principle, the uniform distribution indicates the maximum entropy. So
for each $j$, if it has a size $s_j$, which is almost surely
$O(\log^{a+1}n)$ by Theorem \ref{thm:fundtheorem} (1)(ii), then
$$L_j \leq \log_2 s_j=O(\log\log n),$$
and by average, $$\sum\limits_{j=1}^L \frac{V_j}{2m} L_j=O(\log\log
n).$$

Since by Theorem \ref{thm:fundtheorem} (1)(i), $L$ is almost surely
at most $2n/\log^a n$,
$$L_\mathcal{P} \leq \log_2 L \leq \log n.$$

Note that the number of global edges is $d\cdot L$, which is almost
surely at most $2dn/\log^a n$. Combining them together, we have
$$\frac{m_g}{m} L_\mathcal{P}=O\left(\frac{1}{\log^a n}\right).$$

Thus,
$$L^{\mathcal{P}}(G)= O(\log\log n).$$
The entropy community structure ratio of $G$ by $\mathcal{P}$
$$\tau^{\mathcal{P}}(G)=1-\frac{L^{\mathcal{P}}(G)}{L^U(G)}=1-O\left(\frac{\log\log n}{\log n}\right)=1-o(1).$$

The entropy community structure ratio of $G$
$$\tau (G)=\max_{\mathcal{P}}\{\tau^{\mathcal{P}}(G)\}=1-o(1).$$

(4) follows.

This completes the proof of Theorem \ref{thm:comprinciple}.
\end{proof}

\section{Combinatorial Characteristics Principle} \label{sec:comcharp}

In this section, we prove the combinatorial characteristics principles of homophyly networks, including Theorems \ref{thm:deprinciple}, \ref{thm:widthsp}, \ref{thm:in-in-p}, and \ref{thm:kingp}.

\subsection{Degree Priority Principle}\label{subsec:deg}

\begin{proof} (Proof of Theorem \ref{thm:deprinciple})
Let $T_3=\frac{n}{\log^{a+2} n},
T_4=\left(1-\frac{1}{\log^{(a-1)/2}n}\right)n$.

We just need
to consider the nodes in the homochromatic sets that appear in time
interval $[T_3,T_4]$. We will show that they satisfy the properties
(1)-(4) with probability $1-o(1)$.

\bigskip
For (1) and (2), since for each node $v$, with probability $O(1/t^2)$, there
are at least two edges associated with a newly created seed node
connecting to $v$, the second degree of $v$ is at most one with
probability $1-o(1)$. So the first degree of $v$ is the number of
neighbors of the same color as $v$.
Both (1) and (2) follow.

\bigskip
For (3), note that for a node $v$ of degree $d_v$ at time $t$, the
probability that there is a new seed connecting to $v$ is at most
$\frac{1}{\log^a t}\cdot\frac{d_v}{2dt}\cdot d=\frac{d_v}{2t\log^a
t}=O(\frac{1}{t})$.

Thus the length of degrees of $v$ is expected to
be $O(\log n)$, and so with probability $1-o(1)$, it is at most
$O(\log n)$.

For a node $x$ created after time step $r$, the length of degrees of $x$ is expected to be
bounded by $O(\log\log n)$, so that with probability $1-o(1)$, it is at most $O(\log\log n)$.

\bigskip
For (4), note that a homochromatic set is constructed by
preferential attachment scheme, in which the degree of the first
node is lower bounded by square root of the number of nodes. Since
the size of the homochromatic set is $\Omega(\log^{(a+1)/2})$, the
degree of the seed node contributed by the nodes of the same color,
that is the first degree, is lower bounded by
$\Omega(\log^{(a+1)/4})$.
Theorem \ref{thm:deprinciple} is proved.
\end{proof}

We notice that for homophyly networks, we have only the upper bound of lengths of degrees of nodes.
In applications, both upper and lower bounds of lengths of degrees of nodes may play essential roles.

\subsection{Widths Principle}\label{subsec:widths}

\begin{proof} (Proof of Theorem \ref{thm:widthsp}) Let $N$ be the number of seed nodes in $G$. By Theorem \ref{thm:fundtheorem}, $N=\Omega (\frac{n}{\log^an})$.

Suppose that $t_1< t_2<\cdots<t_N$ are the time steps at which the seed nodes $x_1, x_2,\cdots,x_N$ are created. For each $j$, let
$X_j$ be the community of $x_j$.

By the construction of $G$, we have that for a fixed $j$, for every $t$ with $t_j<t<t_{j+1}$, the node created at time step $t$ contributes to the volume of
a randomly and uniformly chosen community $X_i$ among $X_1, X_2, \cdots, X_j$. Therefore in the interval $(t_j,t_{j+1})$, the volumes of $X_1, X_2\cdots, X_j$
increased uniformly and randomly. At time step $t_{j+1}$, the new seed node $x_{j+1}$ is created. By the construction of $G$, for each $i\in\{1, 2, \cdots, j\}$, the contribution of both widths and volume of $X_i$
is proportional to the volume of $X_i$ immediately before time step $t_{j+1}$. By neglecting the contribution of volumes by global edges, we have that the expected increment of widths of $X_i$ during time step
$t_{j+1}$ is $\Omega (\frac{1}{j})$.

By using the above analysis, we prove our theorem.

For (1), for $X=X_i$ for some $i\leq l$, then $w^G(X)$ is at least $\Omega(\sum\limits_{j=l}^N\frac{1}{j})=\Omega (\log N)=\Omega (\log n)$. (2) follows similarly.
For (3), for an $X=X_i$ for some $i>r$, we have that the width of $X$ is at most $O(\sum\limits_{j=r}^N\frac{1}{j})=O (\log\log N)=O (\log\log n)$. By the choice of $r$, (4) follows from (3).
Theorem \ref{thm:widthsp} follows.
\end{proof}

\subsection{Inclusion and Infection Principle}\label{subsec:in-in-p}

\begin{proof} (Proof of Theorem \ref{thm:in-in-p})
For (1). Let $x$ be a non-seed node created at time step $s$. Then at step $s$, $x$ links only to nodes of the same color. By the construction, for any $t>s$,
if a non-seed node $y$ is created at step $t$, then $y$ has edge with $x$ only if $y$ shares the same color with $x$. (1) holds.

For (2). Let $x$ be a seed node created at time step $s$. Then there are at most $d$ non-seed nodes which link to $x$ during step $s$. By the construction, for any $t>s$,
if a non-seed node $y$ is created at step $t$, then there is no edge between $x$ and $y$ that can be created in step $t$. Therefore $w^G(x)=O(1)$. (2) holds.
Theorem \ref{thm:in-in-p} holds.
\end{proof}

\subsection{King Node Principle}\label{subsec:kingp}

\begin{proof} (Proof of Theorem \ref{thm:kingp})
Suppose that $x_0, x_1,\cdots, x_N$ are all nodes of $X$, created at time steps $t_0<t_1<\cdots<t_N$ respectively.
We use $d(i)[t]$ to denote the degree of $x_i$ at the end of time step $t$. By the construction of $G$, we have that

$$d(0)[t_0]=d,\  d(0)[t_1]\geq 2d,$$

$$ d(i)[t_i]=d$$
\noindent for all $i>0$.

By the construction of $G$, at every time step $t+1$ with $t_i<t+1\leq t_{i+1}$, there is a fixed number $\alpha_t\geq 1$ such that for every $j\leq i$, the expectation of the degree of $x_j$ is amplified by $\alpha_t$.
Therefore $E[d(0)]$ is at least twice of $E[d(i)]$ for all $i\in\{1, 2, \cdots, N\}$. The theorem holds.
\end{proof}

We remark that Theorem \ref{thm:kingp} gives us only some statistical properties of the remarkable role of the seed nodes. Rigorous proofs of the roles of the king nodes need concentration results of the king amplifier,
defined by $\theta (x_0)=\frac{d(x_0)}{\max\{d(x)\ |\ x\in X,\ x\not=x_0\}}$, where $X$ is a community, $x_0$ is the seed node of $X$, for which new methods are needed.

\section{Predicting Principle in Networks } \label{sec:comm}

Theorems \ref{thm:fundtheorem}, \ref{thm:comprinciple}, \ref{thm:deprinciple}, \ref{thm:widthsp}, \ref{thm:in-in-p} and \ref{thm:kingp} show that
there is a structural theory for the homophyly networks. Equally important, our homophyly model explores  that there is a semantical interpretation
for each of the natural communities, that is, nodes of
the same community share common features. We will show that this property provides a principle for
predicting in networks.

To verify that communities of a network are interpretable, we introduce a community finding algorithm.

\subsection{A Community Finding Algorithm}\label{subsec:comfinding}

We design our algorithm by modifying the personalized PageRank vector.
For this, we first review some key ingredients of the PageRank vector and the
related partitioning algorithm which will be useful for us. Given a
graph $G$, an initial vector $s$ on the vertex set and a {\it
teleportation  parameter} $\kappa$, the PageRank vector ${\rm
pr}_\kappa(s)$ is defined recursively as
\begin{eqnarray}
{\rm pr}_\kappa(s)=\kappa s + (1-\kappa){\rm pr}_\kappa(s)W,
\label{eqn:pagerank}
\end{eqnarray}
where $W=\frac{I+D^{-1}A}{2}$ is the lazy random walk on $G$ and $I,
D, A$ denote the identity matrix, the diagonal degree matrix of $G$
and the adjacency matrix of $G$, respectively. It is easy to see
that equation~(\ref{eqn:pagerank}) has a unique
solution. When $s$ equals the indicator
vector $\chi_v$ of vertex $v$, we say that the ${\rm
pr}_\kappa(\chi_v)$ is the \textit{personalized} PageRank vector
with starting vertex $v$ and teleportation parameter $\kappa$. For
an arbitrarily small constant $\epsilon$, there is
an efficient $\epsilon$-approximation algorithm ${\rm
ApproximatePR}(v, \kappa, \epsilon)$ to compute ${\rm
pr}_\kappa(\chi_v)$, which outputs a vector $p={\rm
pr}_\kappa(\chi_v-r)$ \cite{ACL2006a}.

Theorem \ref{thm:fundtheorem} (1) shows that an interpretable community has size bounded by $O(\ln ^{\gamma}n)$, and Theorem \ref{thm:comprinciple} (1) shows that
the conductance of an interpretable community $X$ is bounded by $O(\frac{1}{{{X}}^{\beta}})$ for some constant $\beta$. We will design our algorithm by using these conditions
as the stopping conditions in a personalized pagerank searching. We use $\mathcal{A}$ to denote the
algorithm, which proceeds as follows:

{\bf Algorithm  $\mathcal{A}$}

Given a node $v$ and two constants $\alpha,\beta$, we describe the
algorithm to find a community, if any, as follows: 1) Choose small
constants $\kappa$ and $\epsilon$, and obtain an
$\epsilon$-approximation vector $p$ of the personalized PageRank
vector starting from $v$ with teleportation parameter $\kappa$ by
invoking $ApproximatePR(v, \kappa, \epsilon)$. 2) Do a
\textit{sweep} operation over the vector $p$ such that:
\begin{eqnarray}
\frac{p_{v_1}}{\deg(v_1)}\geq\frac{p_{v_2}}{\deg(v_2)}\geq\cdots\geq
\frac{p_{v_s}}{\deg(v_s)},\label{eqn:order}
\end{eqnarray}
where $s=|supp(p)|$ is the size of the support set of $p$. 3) In
increasing order, for each $i=1,\cdots, s$,  calculate the
conductance of the vertex set $S_i=\{v_1,\cdots,v_i\}$ and output
the {\it first set}, $S_{i}$, satisfying  simultaneously the
following two conditions:
\begin{eqnarray}
\Phi(S_{i})&\leq &\alpha/|i|^\beta, ~\label{eqn:abcondition}\\
\Phi(S_i)&<&\Phi(S_{i+1}).~\label{eqn:nondecrcondition}
\end{eqnarray}

Therefore, our algorithm $\mathcal{A}$ is the approximation algorithm in \cite{ACL2006a} with a new
terminating condition predicted by Theorem \ref{thm:comprinciple} (1).

\subsection{\bf Finding Missing Keywords of Papers from Citation
Networks}

We study Arxiv HEP-TH (high energy physics theory). It is a citation
graph from the e-print arXiv which covers all the citations within a
dataset of $27,770$ papers with $352,807$ edges. If paper $i$ cites
paper $j$, then the graph contains a directed edge from $i$ to $j$.
Each of the papers in the network contains a title, abstract,
publication journal, and publication date of the paper. There are
$1214$ papers among the total $27400$ papers for which keywords were
listed by their authors. We call a paper annotated, if the keywords of the paper have been listed by its authors, and un-annotated, otherwise.
Our goal is to use the annotated papers to
predict and confirm keywords for the un-annotated papers in the
network.

Our homophyly model implies that a community has a short list of
keywords which very well represent the features of the community.

Let $C$ be a community found by our algorithm $\mathcal{A}$ in Subsection \ref{subsec:comfinding}. For some small
constant $i$, we use the most popular $i$ keywords appeared in the
annotated papers in $C$ to represent the common features of $C$,
written ${\rm CF}(C)$. Then we predict that each keyword in ${\rm
CF}(C)$ is a keyword of an un-annotated paper in $C$.

For a keyword $K\in {\rm CF}(C)$, and a paper $P\in C$, we say that
 $K$ is confirmed to be a keyword of $P$, if $K$ appears in either
the title or the abstract of paper $P$.

For each community, $C$ say, suppose that $K_1, K_2,\cdots, K_l$ are
all known keywords among annotated papers in the community $C$. We use the
known keywords $K_1, K_2,\cdots, K_l$ to predict and confirm
keywords for un-annotated papers in $C$. We proceed as follows:

\begin{enumerate}
\item Let $i\leq l$ be a number.
\item Suppose that $K_1, K_2,\cdots, K_i$ are the most popular $i$
keywords among all the known keywords of annotated papers in $C$.
\item Given a un-annotated paper $P$ in $C$, for each $j\leq i$, if $K_j$ appears in either the
title or the abstract of paper $P$, then we say that $K_j$ is a
predicted and confirmed keyword of paper $P$.

\end{enumerate}

\begin{figure}
\centering
\includegraphics[width=3in]{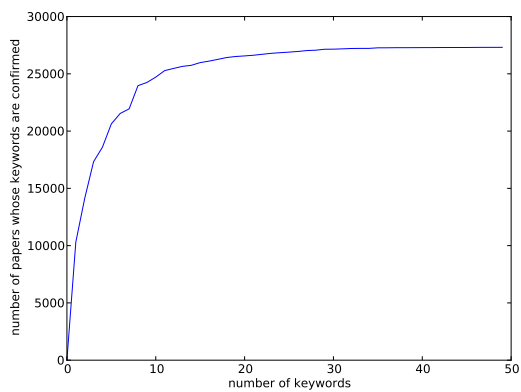}
\caption{Keywords prediction. The curve represents the numbers of
papers whose keywords are predicted and confirmed by using the most
popular $k$ keywords as the common keywords of all the communities,
for $k\leq 50$. The curve increases quickly and becomes flatten
after $k>10$. This means that each community has a few ($10$)
remarkable common keywords, a result predicted by the homophyly
networks.} \label{figure_keywords prediction 1}
\end{figure}

In Figure \ref{figure_keywords prediction 1}, we depict the curve of numbers of papers whose keywords
are predicted and confirmed for $i$ up to $50$, where $i$ is the number with which the most popular $i$ keywords
are used to predict the keywords of the community, for all the communities.

The results in Figure \ref{figure_keywords prediction 1} show that a community of the citation network can
be interpreted by the most popular $10$ keywords and that the
interpretations of communities can be used in predictions and
confirmations of functions of nodes in the network. This experiment shows that for each
community, nodes of the same community do share common features,
that is, the short list of common keywords, and that the common
features of each of the communities can be used in predicting and
confirming functions in networks. Our homophyly model predicts that this
property may be universal for many real networks. This provides a principle for predicting and confirming functions in
networks.

\subsection{Homophyly Law of Networks}

In Table \ref{tab:keywords-p}, we describe the full prediction and confirmation of keywords of papers in the network.
In the table, the first row shows that if we define the keywords of
a community to be the most popular $5$ keywords of annotated papers in
the community, then the prediction algorithm above finds exactly $1$
keyword for $5286$ papers, $2$ keywords for $2979$ papers, $3$
keywords for $1639$ papers, $4$ keywords for $768$ papers, $5$
keywords for $345$ papers, $6$ keywords for $166$ papers, $7$
keywords for $65$ papers, $8$ keywords for $21$ papers, $9$ keywords
for $7$ papers, and even $10$ keywords for $2$ papers. In total,
there are $11279$ papers to each of them there is at least one
keyword is predicted and confirmed. The second row shows the number
of papers for which $r$ keywords are predicted and confirmed for all
$r\in [1,10]$, in the case that we define the keywords of a community to be the
most popular $10$ known keywords of annotated papers in the community. In this case,
there are $13795$ papers in total each of which has at least one
keyword is predicted and confirmed. Table \ref{tab:keywords-p} shows that most communities have a short list
of representative keywords, this is $10$ or even $5$. This is exactly the result predicted by our homophyly model.

\begin{center}
\begin{table*}[ht]

\begin{tabular} {|c|c|c|c|c|c|c|c|c|c|c|c|c|c|c|c|c|}
\hline {$i$}$\backslash${$r$}
\makebox&1&2&3&4&5&6&7&8&9&10&total\\
\hline
5&5286&2979&1639&768&345&166&65&21&7&2&11279\\
\hline
10&4701&3605&2429&1407&790&434&236&102&48&23&13795\\
\hline
15&4360&3627&2671&1798&1074&606&340&178&95&37&14829\\
\hline
20&3953&3467&2853&1999&1310&798&462&268&144&67&15397\\
\hline
25&3666&3301&2909&2116&1498&912&575&342&201&81&15721\\
\hline
30&3344&3169&2934&2223&1648&1053&692&410&253&129&16015\\
\hline
35&3199&3116&2952&2238&1681&1152&728&460&272&145&16151\\
\hline
40&3081&3044&2922&2255&1734&1218&752&500&323&158&16239\\
\hline
45&2987&2992&2850&2321&1741&1253&836&517&364&181&16333\\
\hline
50&2869&2915&2770&2340&1844&1291&880&579&413&214&16453\\
\hline
all&2336&2587&2560&2348&1883&1528&1150&849&568&373&16842\\
\hline
\end{tabular}
\caption{Keywords prediction}\label{tab:keywords-p}
\end{table*}
\end{center}

Let $G=(V,E)$ be a network. Suppose that each node $v\in V$ has some colors. For a node $v\in V$, we use $D(v)$ to denote the number of colors
associated with $v$. We define the {\it dimension of $G$} by

$${\it dim}(G)=\max\limits_{v\in V}\{D(v)\}.$$

Suppose as usual that in a citation network, $G$ say, each paper has up to $5$ keywords. Then $G$ has dimension $5$.
By Figure \ref{figure_keywords prediction 1} and Table \ref{tab:keywords-p},
we observe the following property: Given a network $G$, if $G$ has dimension $D$, then, each community, $C$ say, of $G$ can be interpreted by
$O(D)$ many common colors of the community $C$. This experiment, together with our homophyly model, predicts that the following predicting principle may hold
for many real networks.

 {\it Predicting principle}: Let $G$ be a network of dimension $D$. Then for a (typical or natural) community $C$ of network $G$, there is
 a list of functions of length $O(D)$ which represent the common features of nodes in $C$, so that $C$ is interpreted by a list of common features as short as
$O({\it dim}(G))$.

This principle provides not only the mechanism for network predicting, but also a quantitative criterion for predicting functions in networks.

\section{C-Community Structure Ratio of Real Networks}\label{sec:conreal}

In Definition \ref{def:algorithm}, we defined the conductance community structure ratio of a network given by an algorithm, $\mathcal{A}$ say.
This suggests the algorithmic problem to find the algorithm
which finds the maximal conductance community structure ratios of
networks.

In Section \ref{sec:real}, we used three algorithms
$\mathcal{C}$, $\mathcal{E}$ and $\mathcal{M}$ based on conductance, entropy and modularity definitions of community structures respectively, where $\mathcal{C}$ is the algorithm $\mathcal{A}$ in Subsection \ref{subsec:comfinding}.
Here we use these algorithms again to compute the conductance community structure ratios of the networks given by the three algorithms.
In Table \ref{table_modularity}, we report the conductance community
structure ratios of the algorithms $\mathcal{C}$, $\mathcal{E}$ and
$\mathcal{M}$ on $22$ real networks.

\begin{table}
\centering
\begin{tabular} {|c|c|c|c|}
\hline
Networks$\backslash$Algorithms&$\theta^{\mathcal{C}}(G)$&$\theta^{\mathcal{E}}(G)$&$\theta^{\mathcal{M}}(G)$\\
\hline
football&0.97&0.76&0.74\\
\hline
cit-hepph&0.7&0.83&0.19\\
\hline
cit-hepth&0.59&0.54&0.31\\
\hline
col-astroph&0.72&0.56&0.25\\
\hline
col-condmat&0.84&0.55&0.77\\
\hline
col-grqc&0.96&0.72&0.82\\
\hline
col-hepph&0.77&0.8&0.24\\
\hline
col-hepth&0.89&0.67&0.7\\
\hline
p2p24&0.83&0.46&0.51\\
\hline
p2p25&0.85&0.56&0.54\\
\hline
p2p30&0.84&0.58&0.5\\
\hline
p2p31&0.82&0.52&0.54\\
\hline
p2p4&0.87&0.6&0.38\\
\hline
p2p5&0.91&0.71&0.4\\
\hline
p2p6&0.92&0.56&0.37\\
\hline
p2p8&0.94&0.81&0.47\\
\hline
p2p9&0.92&0.80&0.46\\
\hline
email-enron&0.73&0.55&0.48\\
\hline
email-euall&0.77&0.85&0.25\\
\hline
road-ca&0.98&0.92&0.996\\
\hline
road-pa&0.98&0.97&0.99\\
\hline
road-tx&0.99&0.94&0.99\\
\hline
\end{tabular}
\caption{conductance community structure ratios of community detecting
algorithms based on minimal conductance, information flow and
modularity on some real networks, written by $\mathcal{C}$,
$\mathcal{E}$ and $\mathcal{M}$ respectively.}
\label{table_modularity}
\end{table}

Table \ref{table_modularity}
shows that for most real networks, algorithm $\mathcal{C}$ finds the
largest conductance community structure ratio, and that for some real
networks, algorithm $\mathcal{E}$ finds the largest
conductance community structure ratio. A common property of all the
real networks is that the conductance community structure ratios of all
real networks are large. In fact, there are $6$ networks whose
conductance community structure ratios are greater than $0.9$, there
are $9$ networks whose conductance community structure ratios are
between $0.8$ and $0.9$, there is one network whose
conductance community structure ratio is between $0.7$ and $0.8$, and
there is one network whose conductance community structure ratio is at
least $0.59$. These results show that each of the real networks has
a remarkable community structure. However it is hard to have a single algorithm which finds the maximal conductance community structure ratios $\theta$'s for all networks.

\section{Test of Community Finding Algorithms}\label{sec:test}

Given a network, $G$ say, and a community finding algorithm
$\mathcal{A}$, we have a conductance community structure ratio
$\theta^{\mathcal{A}}(G)$.

Theoretically speaking, for two algorithms $\mathcal{A}$ and $\mathcal{B}$, if
$\theta^{\mathcal{A}}(G)>\theta^{\mathcal{B}}(G)$, then $\mathcal{A}$ is better than $\mathcal{B}$ for $G$.
However, we don't know: what does this mean in real networks analyses and real world applications?

We use the three algorithms $\mathcal{C}$, $\mathcal{E}$, and $\mathcal{M}$ in Section \ref{sec:conreal} again.
We implement the
keywords prediction and confirmation on the same citation network
based on three community finding algorithms $\mathcal{C}$, $\mathcal{E}$, and $\mathcal{M}$ respectively. We found that the conductance community
structure ratios of the network $G$ given by $\mathcal{C}$, $\mathcal{E}$, and
$\mathcal{M}$ are $0.59$, $0.54$ and $0.31$ respectively (referred to Table \ref{table_modularity}). We depict
the curves of keywords predictions and confirmations of the three
algorithms on the citation network in Figure \ref{figure_keywords
prediction 2}.

\begin{figure}
  \centering
\includegraphics[width=3in]{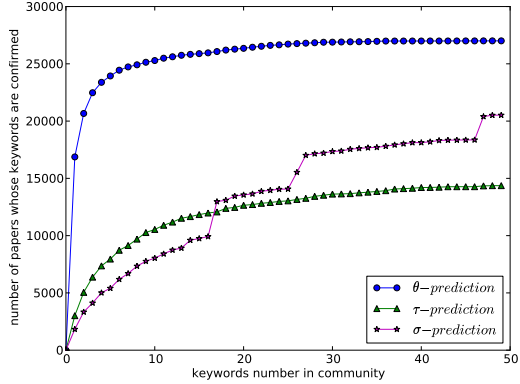}
\caption{Keywords Prediction. The three curves correspond to the
results of prediction and confirmation of the three algorithms
$\mathcal{C}$, $\mathcal{E}$ and $\mathcal{M}$, marked by $\theta$-, $\tau$-,
and $\sigma$-predictions respectively in the figure. This figure shows that
$\theta^{\mathcal{C}}(G)$ is the largest among the three algorithms,
and at the same time, the prediction and confirmation of keywords
based on the communities found by algorithm $\mathcal{C}$ are the
best among that of the three algorithms.}
 \label{figure_keywords prediction 2}
\end{figure}

From Figure \ref{figure_keywords prediction 2}, we observe that algorithm $\mathcal{C}$ has the largest conductance
community structure ratio and the best
performance of keywords prediction and confirmation.
This means that larger conductance community structure ratio implies a better interpretation of communities and a better
performance in predicting and confirming functions in networks. Therefore maximizing conductance community structure ratios of networks
does have implications in real world network analyses and applications.

\section{Conclusions and New Directions}
\label{sec:newissues}

We proposed a new model of networks, the homophyly model, based on which
we built a structural theory of networks. Our theory is a mathematical understanding of networks. However, the homophyly model is motivated by
observing the connecting behaviors in nature and society, therefore the high level open issue is to explore the social, biological and physical understandings of the homophyly model.

The fundamental results of our theory are: community structures are definable, and
communities are interpretable in networks. The two results point out the syntax and semantic aspects of networks respectively. We believe that our research provides a firm foundation for
a structural theory of networks. However, to fully develop such a theory, there are a number of new issues left open. We discuss a few of the most important ones here.

 The first is
a non-linear or high dimensional network theory. Given a network, $G$ say, in which each node has some colors, for each node $v$, we use $D(v)$ to denote the number of colors
associated with it. In this case, we define the dimension of $G$ to be the maximal of $D(v)$'s among all nodes $v$, that is, ${\it dim}(G)=\max\limits_{v\in V}\{D(v)\}$. By this definition, our homophyly networks all have dimension $1$, so that they are linear
networks. Therefore our theory is a linear network theory. Clearly, it is interesting to develop a non-linear or high dimensional network theory, which is expected to be harder, since there would be more combinatorics
involved in the theory.

The second is a global theory of networks. We regard communities as local structures of networks. This means that our theory is a local theory of networks, predicting a global theory of networks simultaneously.

The third is to develop new theory and applications based on the principles of the community structures discovered in the present paper. For this, we introduce a few of them:

\begin{enumerate}

\item [(1)] To understand the nature and to develop applications of the holographic law predicted here in large-scale real network data

Our theory predicts that for a large-scale real network, $G$ say, the power exponent of the power law of $G$ is the same as that of a natural or typical community
$G_X$ of $G$ for some set $X$ of size polynomial in $\log n$, where $G_X$ is the induced subgraph of $X$ in $G$. This would be an interesting new phenomenon of real world big data. New
applications of the result are of course possible. For instance, in network searching, we may find a community $X$ of size as large as a few hundreds, which is still large for real recommendation. By
the holographic law, there is a small set, $X_0\subset X$ which almost dominates $X$, in which case, $X_0$ could be as small as $10$ to $20$ nodes. In so doing, we could simply recommend $X_0$, which would keep
the most useful information of $X$.

\item [(2)] To understand the roles of external centrality of communities of a network

Our widths principle predicts that for a natural community $X$ of a network $G$ such that the size of $X$ is polynomial in $\log n$, where $n$ is the number of nodes in $G$, there is a set $X_0\subset X$ such that
$X_0$ is as small as $O(\log\log n)$ and such that $X_0$ almost dominates the external links from $X$ to outside of $X$. This property is useful in both theory and applications. For instance, in a citation network $G$,
we may find a community of $G$, $X$ say, in which case, $X$ could be interpreted as the papers on some topic, and the external centrality set $X_0$ of $X$ could be interpreted as the papers having influence on research of other topics.
By extracting the keywords of papers in $X_0$, we may already know much of the relationships between the topic of $X$ and the topics relevant to that of $X$.

\item [(3)] To understand the roles of the local communication law in network communications

Our fundamental theorem says that the diameters of natural communities are bounded by $O(\log\log n)$. This means that in a communication network, the most frequent communications are local ones which are
exponentially shorter than that of a global communication, and that global communications are much less frequent. This provides an insight to analyze the complexity of communications in networks.

\item [(4)] To investigate new notions of networks that are locally collective by using our local theory of networks

We understand that communities of a network are local structures of the network, and that there are important notions of networks which are locally collective. Our theory provides an insight to
study the locally collective notions by the community structure of networks. Here we discuss one of the most important such notions, the happy node problem below. In a homophyly network, $G$ say,
each node $v$ is associated with a color. We could define happiness as follows: we say that a node $v$ is happy in $G$, if all the neighbors of $v$ share the same color as $v$. With this definition, we know that
most nodes are happy in $G$. On the other hand, the diameter of a community in $G$ is $O(\log\log n)$. This means that for a node $v$, whether or not $v$ is happy in $G$, is independent of nodes $\Omega (\log\log n)$ far away from $v$.
These observations provide an insight to build happiness of individuals as a locally collective notion of networks, which calls for further investigation.

\item [(5)] To develop a security theory of networks based on the structural theory of networks

The first achievement of this is the security model and provable security of networks in \cite{LPZ2013a}, in which a number of open problems were posed.

\item [(6)] To develop a theory of evolutionary games in networks based on the structural theory of networks

This is possible by our work in \cite{LPY2013a}. The goal of this theory would solve some long standing challenges in social science and economics. The later mission is of course a grand challenge.

\item [(7)] Approximation and hardness of approximation of the conductance community structure ratio of networks

Our definition of the conductance community structure ratio provides a way to test the quality of community finding algorithms. In real networks, each of the algorithms based on personalized pagerank,
compression of information flow and modularity has reasonably good performance in finding the conductance community structure ratios. However, it is an open question to prove some theoretical results
for approximation and hardness of approximation of the problem.

\item [(8)] To prove theoretically that the community structure hypothesis holds for networks of other classical models such as the ER and PA models.

We have shown experimentally that the hypothesis holds for networks of both the ER and PA models. It would be interesting to have theoretical proofs of the results. Generally, it is interesting to prove the hypothesis for
networks of all reasonable models.

\end{enumerate}

\bibliographystyle{plain}
\bibliography{acmsmall-homophyly-2013-2-bibfile}

\end{document}